\documentclass[11pt,a4paper,twocolumn]{article} 

\usepackage[a4paper,top=2cm,bottom=2cm,left=2.5cm,right=2.5cm,marginparwidth=1.75cm]{geometry}

\usepackage[english]{babel}
\usepackage{lineno} 

\rightlinenumbers 

\usepackage[square,numbers,sort&compress]{natbib}
\usepackage{chapterbib}
\usepackage{multicol}
\usepackage{amsmath}
\usepackage{amssymb}
\usepackage{graphicx}
\usepackage[colorlinks=true, allcolors=blue]{hyperref}
\usepackage{hyperref}
\usepackage{bookmark}
\usepackage[title]{appendix}
\usepackage{mathrsfs}
\usepackage{amsfonts}
\usepackage{booktabs} 
\usepackage[format=plain,
            labelfont={bf,sl}]{caption}
\usepackage{threeparttable} 
\usepackage{algorithm}
\usepackage{algorithmicx}
\usepackage{algpseudocode}
\usepackage{listings}
\usepackage{enumitem}
\setlist[itemize]{noitemsep}
\usepackage{chngcntr}
\usepackage{booktabs}
\usepackage{lipsum}
\usepackage{subcaption}
\usepackage{authblk}
\usepackage[T1]{fontenc}    
\usepackage{csquotes}       
\usepackage{diagbox}
\usepackage{xcolor}
\usepackage{textcomp}
\usepackage{gensymb}

\usepackage{siunitx}
\usepackage{hhline}

\usepackage{setspace}

\usepackage{titlesec}
\titleformat{\section} 
  {\normalfont\Large\bfseries}{\thesection.}{1em}{}

\newcommand{\insnte}{In$_{x}$Sn$_{1-x}$Te }

\title{Induced superconductivity in selective-area grown SnTe devices}

\author[1]{Maarten J.G. Kamphuis\textsuperscript{\textdagger}}
\author[2]{Yoran F.S. Starmans\textsuperscript{\textdagger}}
\author[2]{Pim J.H. Lueb}
\author[3,1]{Femke J. Witmans}
\author[2]{Marvin M. Jansen-Zilles}
\author[2,4]{Marcel A. Verheijen}
\author[2]{Reinoud Lavrijsen}
\author[1]{Joost Ridderbos}
\author[5,2]{Fabrizio Nichele}
\author[1]{Floris A. Zwanenburg}
\author[2]{Erik P.A.M. Bakkers}
\author[1]{Alexander Brinkman}

\affil[1]{\small MESA+ Institute, University of Twente, The Netherlands}
\affil[2]{\small Department of Applied Physics, Eindhoven University of Technology, 5600 MB Eindhoven, The Netherlands}
\affil[3]{\small II. Physikalisches Institut, Universität zu Köln, Zülpicher Str. 77, D-50937 Köln, Germany}
\affil[4]{\small Eurofins Materials Science Eindhoven, 5656 AE Eindhoven, The Netherlands}
\affil[5]{\small IBM Research Europe Zürich, 8803 Rüschlikon, Switzerland}

\date{}  

\begin{document}
\onecolumn
\maketitle

\begin{abstract}    
Topological superconductors are of high interest for applications in topological quantum computation. The required topologically superconducting state can be engineered by proximity-inducing superconductivity in a topological insulator. Here, we explore the induced superconductivity in selective-area grown nanowires of the topological crystalline insulator SnTe on a InP substrate, through TEM/EDX and low-temperature electronic transport studies. The observed superconducting behavior likely originates from indium in the substrate diffusing upwards into the SnTe nanowire, forming a thin layer of In\textsubscript{x}Sn\textsubscript{1-x}Te at the interface between the nanowire and the substrate. In\textsubscript{x}Sn\textsubscript{1-x}Te is intrinsically superconducting for indium concentrations above 2$\%$, resulting in superconductivity within the heterostructure. Little-Parks oscillations are observed in loop-shaped nanowire networks in an out-of-plane magnetic field. The half-period shift indicative of a topological superconducting state is absent, which is explained by dominant trivial transport channels obscuring any topological signatures.
\end{abstract}

\begin{figure}[h!]
    \centering
    \includegraphics[width=0.7\linewidth]{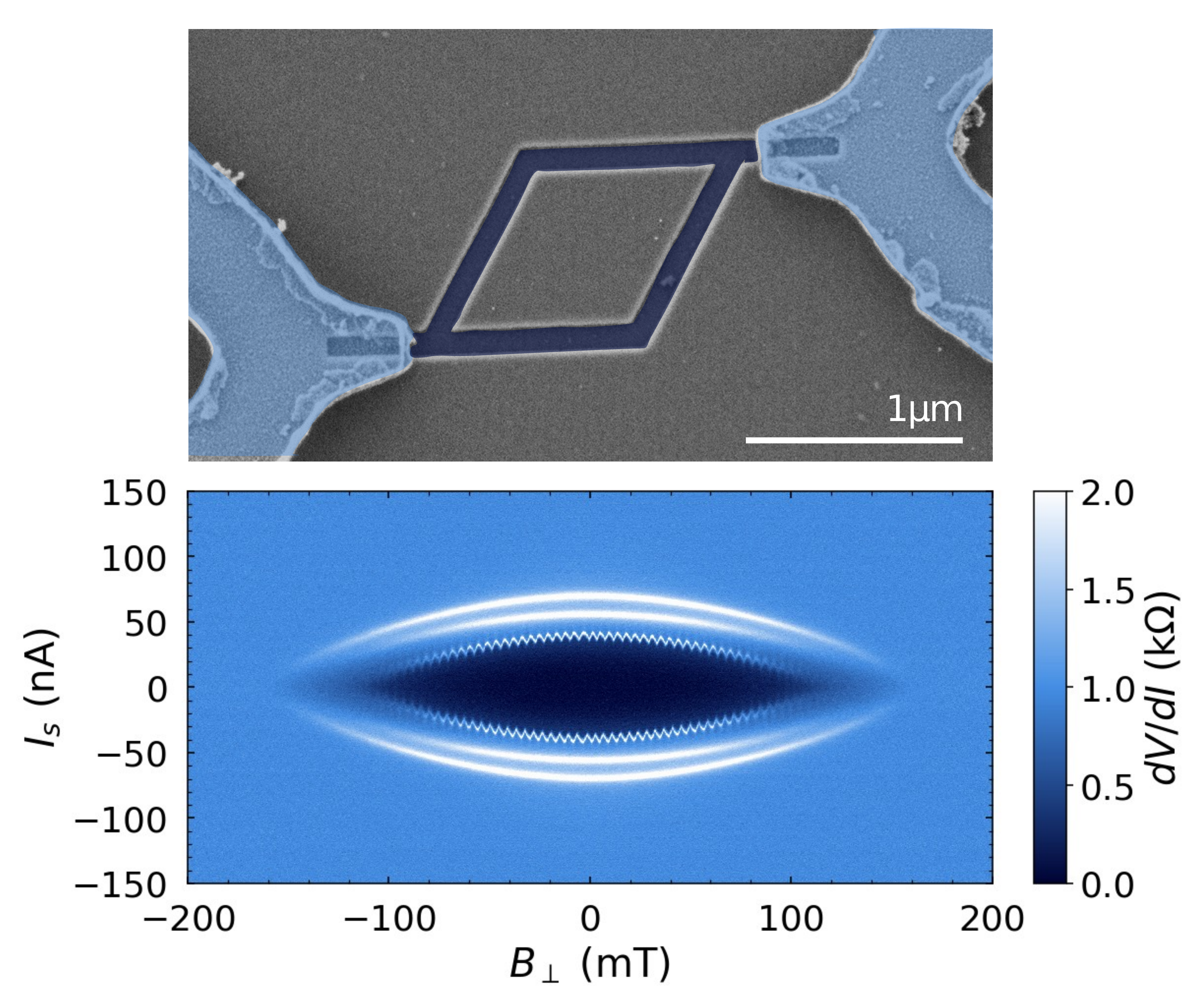}
    \label{fig - abstract}
\end{figure}


\newpage
\twocolumn
Topological insulators (TIs) are bulk insulators with spin-polarized surface states, originating from a band inversion around the band gap. These materials garner significant interest in systems where their hallmark properties are combined with superconductivity. 
The resulting topological superconductor (TSC) is highly sought after, because they are predicted to host emergent zero-energy quasi-particle excitations known as Majorana zero modes~\cite{Qi_Time-reversal_2009, Sato_Topological_2010, Qi_Topological_2011}. These Majoranas, as they are colloquially known, are the foundation for topological quantum computation~\cite{Freedman_Topological_2001, Stern_Topological_2013}. 
Some TIs can become intrinsically superconducting through doping (e.g. intercalated Bi$_2$Se$_3$~\cite{Hor_Superconductivity_2010, Sasaki_Topological_2011}). However, in many instances, the TSC state needs to be engineered in hybrid systems where a TI is coupled to an \textit{s}-wave superconductor~\cite{FuKane_Superconducting_2008, CookFranz_Majorana_2011, CookFranz_Stability_2012, Xu_Artificial_2014}. This idea can be extended to the topological crystalline insulator (TCI) material class, of which tin telluride (SnTe) is the textbook example~\cite{Hsieh_Topological_2012, Tanaka_Observation_2012, Tanaka_Two_2013}. In a TCI the topological surface states (TSS) are protected by crystalline space group symmetries instead of time-reversal symmetry, as is the case in conventional topological insulators~\cite{Fu_Topological_2011}.

Efforts have been made to engineer the TSC phase in SnTe, both by proximity-inducing superconductivity~\cite{Klett_Proximity-induced_2018, Rachmilowitz_Proximity-induced_2019, Trimble_Josephson_2021, Singh_Complex_2023} and by doping with indium. Multiple studies have shown that an indium concentration of just $2\%$ already results in a transition temperature, $T_\text{c}$, of up to several Kelvins~\cite{Balakrishnan_Superconducting_2013, Zhong_Indium_2017} and is found to increase with indium concentration~\cite{Bushmarina_Superconducting_1991, Gonzalez_Superconductivity_2025, Zhong_Optimizing_2013}.
Most experimental studies show \insnte to be a trivial type-II \textit{s}-wave superconductor~\cite{Haldolaarachchige_Anomalous_2016, Denisov_Superconducting_2020, Denisov_High-field_2025, Gorina_Two-band_2017, Smylie_Nodeless_2020}, especially for high indium concentrations. However, low indium concentrations bear more promises for topological superconductivity as pointed out by studies performed on \insnte samples with 4.5$\%$ In: Point-contact spectroscopy has revealed the presence of surface Andreev bound states~\cite{Sasaki_Odd-parity_2012}, while angle-resolved photoemission spectroscopy measurements, though somewhat obscured due to increased scattering from the indium doping, have confirmed the Dirac nature of the surface state~\cite{Sato_Fermiology_2013}.

Here, we report the observation of superconductivity in selective-area grown (SAG) SnTe nanowire devices, epitaxially grown on an InP(111)A substrate with a SiN$_x$ mask. As we will argue, the diffusion of In from the substrate into the SnTe during nanowire growth is the most likely cause of the superconductivity, despite the fact that our spectroscopic analysis enforces an upper bound of a few percent In content in the SnTe at the interface. Little-Parks oscillations observed in loop-shaped devices were used to analyse the nature of the superconductivity. However, no indications of a sign-changing order parameter are observed, suggesting that the observed superconductivity is dominantly present within the trivial bulk states. Nevertheless, our studies provide a novel platform for realizing in-situ deposited topological superconductors once trivial bulk states are depleted.  

\begin{figure}[h!!]
    \centering
    \includegraphics[trim=10pt 0pt 0pt 10pt, clip, width=0.92\linewidth]{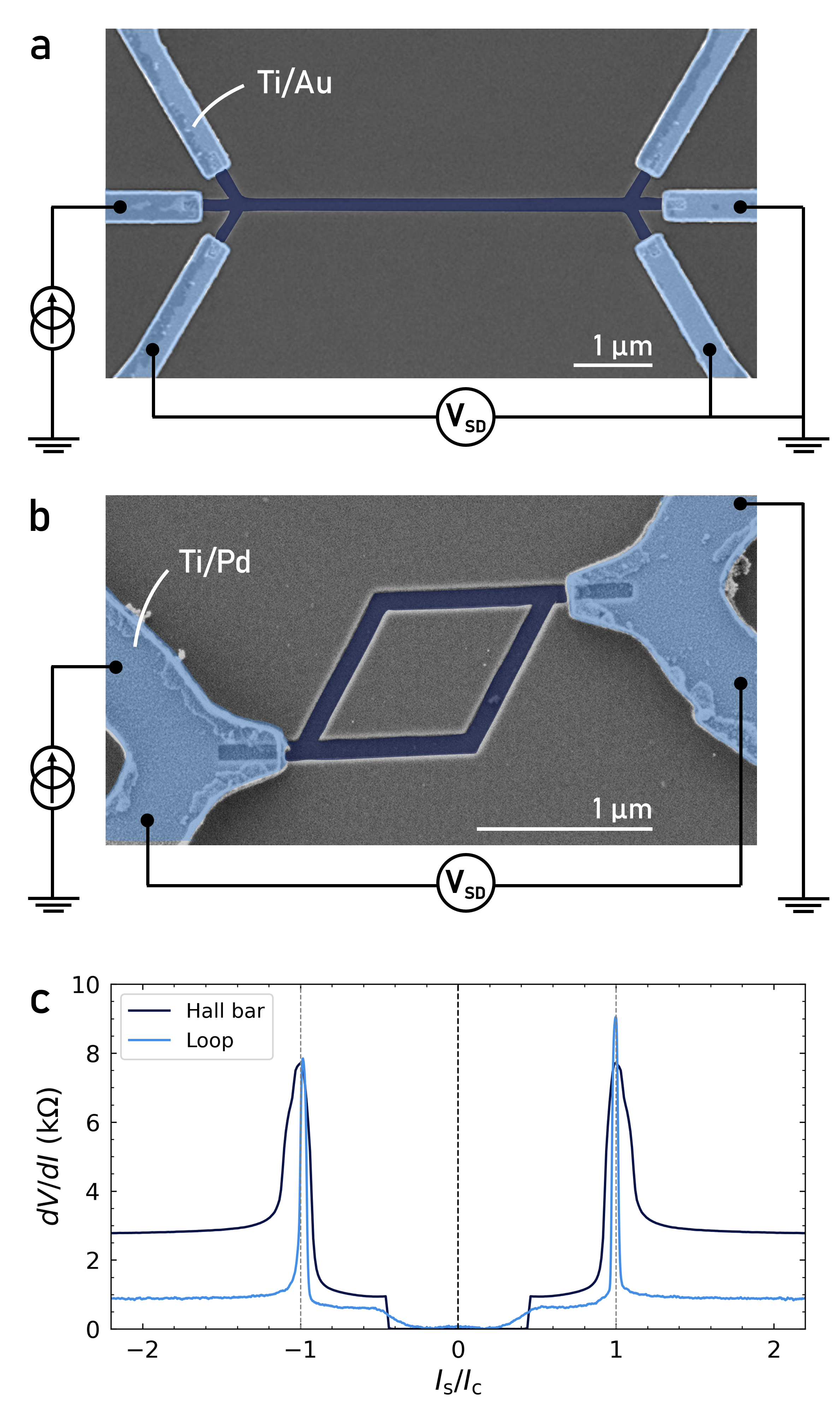}
    \caption{\textbf{Superconductivity in SAG SnTe nanowire devices}. False-colored top-view SEM images of SAG SnTe Hall bar (a) and loop (b) devices. The SnTe nanowires are colored in dark blue, the metal contacts in light blue. Also indicated schematically is the (quasi-)4-point wiring configuration for low-temperature electronic transport measurements. (c) Typical example of the measured differential resistance $dV/dI$ as a function of the normalized applied current $I_\text{s}/I_\text{c}$ for both device types. A re-entrant resistance, indicative of type-II superconductivity, is observed consistently for all devices before the transition to the normal state.}
    \label{fig:SEM+SC}
\end{figure}

The fabricated nanowires have a (111)-oriented top surface, and can be reproducibly grown as narrow as 50~nm over lengths larger than 10 \textmu m. The selective-area growth method allows for the design of more complex nanowire networks, in which the branches follow $\langle110\rangle$ or $\langle112\rangle$ in-plane crystal directions. This enables, for instance, the definition of Hall bars and loops which we will investigate in this work. Specifically, the Hall bar configuration allows to determine the normal state carrier density of the SnTe and the loops allow to study fluxoid quantization in the superconducting state. For more details on the growth of these samples, see  Schellingerhout \textit{et al.}~\cite{Schellingerhout_In-plane_2023}. To form Ohmic contacts on these nanostructures, the native oxide was removed by an Ar-ion etch, directly followed by deposition of the metal contacts. On all Hall bar devices Ti(10nm)/Au(100nm) contacts have been deposited using e-beam evaporation. On all loop devices Ti(1nm)/Pd(30nm) contacts have been deposited by a combination of sputtering and e-beam evaporation. For details, see the methods section. 

False-colored scanning electron microscopy (SEM) images of one of each device type are shown in Figure~\ref{fig:SEM+SC}a and b, respectively. Here we also indicate the wiring for the low-temperature electronic transport measurements. The measured differential resistance $dV/dI$ as a function of the source current is plotted in Figure~\ref{fig:SEM+SC}c for both device types. The observed behaviour is indicative of type-II superconductivity: At low bias, the differential resistance is zero, signalling the superconducting nature of the transport. This state persists up to a critical current $I_\text{c1}$, above which a re-entrant resistance is observed, consistent with the flux-flow regime of a type-II superconductor, where unpinned vortices move and give a voltage. At a second critical current, $I_\text{c2}$, the device switches to the normal state. This behaviour is observed for all 12 measured devices. An additional effect that can lead to a re-entrant resistance is the inverse proximity effect of normal state contact on a superconducting channel. Recent work by Oliveira \textit{et al.} \cite{Oliveira_Anomalous_2025} on this effect closely reproduces some detailed features in the re-entrant resistance regime in our data.

We confidently discard the possibility that the superconductivity is induced by the contacts. The only element that might be superconducting at the temperature of the experiment is Ti ($T_\text{c} <$ 300~mK~\cite{Steele_Superconductivity_1953, Wang_Electron_2018}). However, that would require that a few nanometers of Ti maintain its superconducting properties when covered with 30 nm of Pd or 100 nm of Au, which is exceedingly unlikely. Moreover, if superconductivity is induced throughout the entire length of a nanowire device by Ti-containing contacts on either end, the entire system should behave a s a Josephson junction with the SnTe acting as a weak link. However, no signatures of Josephson-junction behaviour are observed, both in DC and radiation experiments (see Supplementary~\ref{Suppl:Contacts},~\ref{Suppl:Shapiro}). 
Moreover, in some of the devices this scenario would require a weak link exceeding 5~\textmu m in length, which is significantly longer than our estimated value of $\xi_\mathrm{D}\sim160$ nm (for derivation see Appendix~\ref{Suppl:Xi}). This estimated value is in the same order-of-magnitude as the longest reported superconducting coherence length in SnTe in literature: $\sim120$~nm~\cite{Klett_Proximity-induced_2018}. 
Other contact- or capping-related materials (Pd, Au, and AlO\textsubscript{x}) are excluded as a cause for superconductivity, since the two devices employ different material stacks, yet exhibit very similar superconducting behaviour. 

Another potential cause for superconductivity is found in the bulk of the film itself: SnTe is usually strongly p-doped, due to the negative formation energy of tin vacancies~\cite{Wang_Microscopic_2014}. Intriguingly, Sn$_{1-\delta}$Te (with $\delta$ the fraction of vacancies at the tin lattice sites) becomes intrinsically superconducting for sufficiently high carrier densities~\cite{Hulm_Superconducting_1968, Allen_Carrier_1969}.
Hall effect measurements on our samples in the normal state reveal carrier densities of around $2\cdot10^{20}$~cm$^{-3}$ (see Supplementary~\ref{Suppl:Hall}), which approximately corresponds to the reported onset of superconductivity in SnTe bulk crystals in the mentioned studies. However, the critical temperature measured in our devices ($T_\text{c}$ = 350 mK) exceeds their predicted $T_\text{c}$ by two orders of magnitude.
Moreover, reports on electronic transport in SnTe nanostructures typically show similar carrier densities, without any signatures of superconductivity ~\cite{Volobuev_Giant_2017, Shen_Synthesis_2014, Mientjes_Structural_2025}. Therefore, we exclude Sn vacancy doping as a likely source for the observed superconductivity.

Another possible mechanism is interface-induced superconductivity analogous to that investigated for SnTe-PbTe heterostructures~\cite{Tang_Strain_2014, Sidorczak_PbTe_2026}. Although it is unknown whether a similar mechanism exists at the SnTe-InP interface, this appears unlikely in our devices, since the reported critical temperatures (3-6 K) are substantially higher than the observed $T_\text{c}= 350$mK. Additionally, the proposed flat-band mechanism is expected only for carrier densities below $\sim10^{12}$ cm\textsuperscript{-2}, far below those in our devices. We therefore consider a strain-induced mechanism unlikely in our devices.

The remaining plausible scenario is superconductivity arising from indium doping, enabled by indium diffusion from the InP substrate into the SnTe nanowires. 
Indium substitution doping is exothermic~\cite{Gonzalez_Superconductivity_2025}, and has been reported in \insnte/SnTe thin film stacks~\cite{Balakrishnan_Superconducting_2013}. Notably, superconductivity has been reported in \insnte for indium concentrations as low as $x \approx 2\%$, with critical temperatures and critical fields comparable to those observed in our transport measurements~\cite{Balakrishnan_Superconducting_2013, Zhong_Optimizing_2013, Zhong_Indium_2017, Erickson_Enhanced_2009}.

\begin{figure*}[h!]
\includegraphics[width=\linewidth]{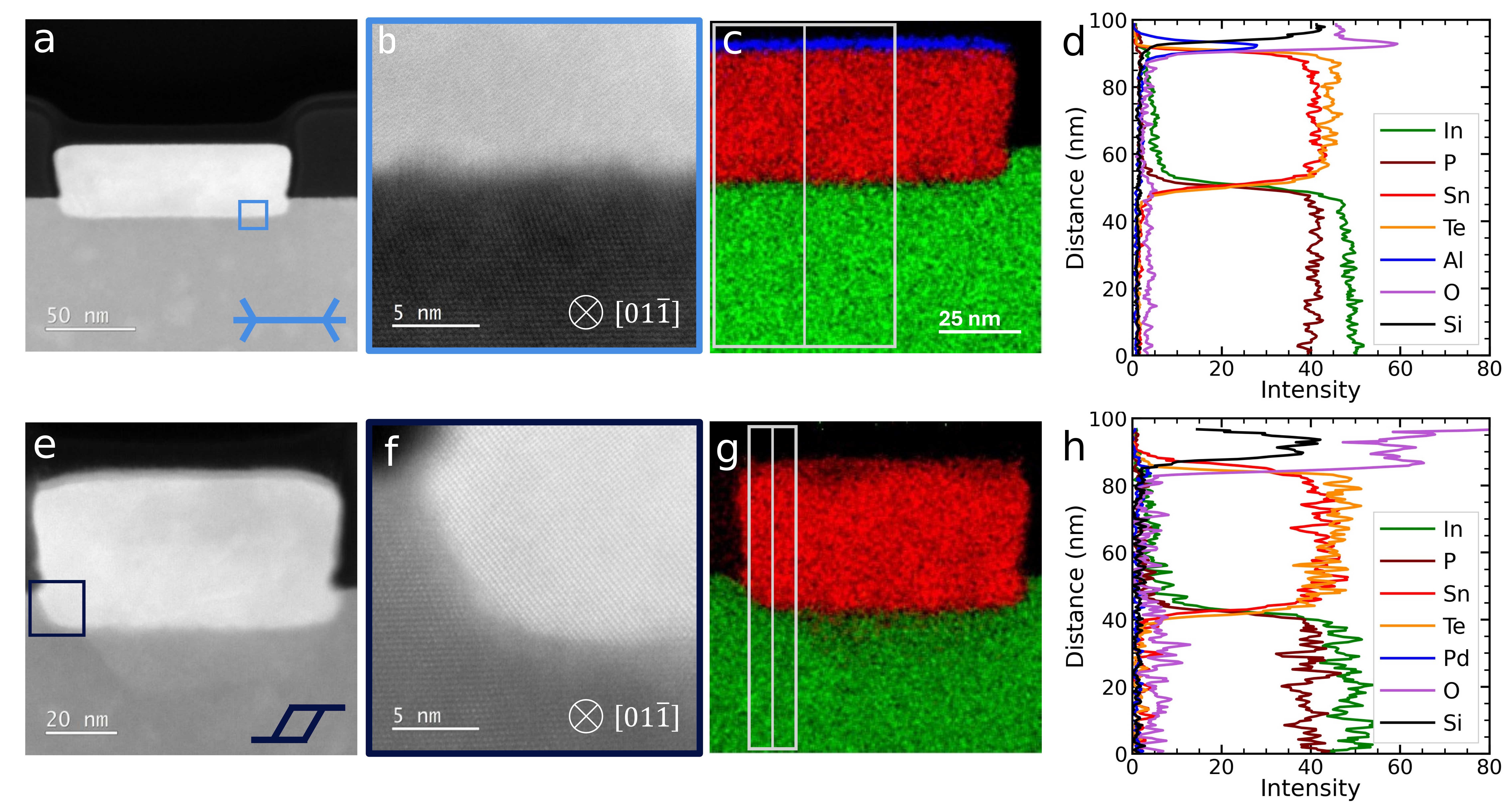}
\caption{\textbf{Cross-sectional HAADF-STEM and EDX measurements on both device types.} (a,e) HAADF-STEM cross section for the Hall-bar and AB-loop devices respectively. (b,f) Atomic resolution HAADF-STEM micrographs of the InP-SnTe interface. (c,g) EDX compositional mappings showing the elemental distributions of In (green), Sn (red) and Al (blue). No significant intermixing of the InP and SnTe constituents can be observed. (d,h) Line profiles extracted from the white boxed area in Figures (c,g). No evidence for indium diffusion can be discerned.}
\label{fig:TEM-EDX}
\end{figure*}

Indium is a valence skipping element and occurs in both the +1 and +3 valence state, and substitutes on Sn lattice sites, resulting in the formation of an impurity band that overlaps with the valence band~\cite{Wang_Evidence_2024}. Both valence species occur in equal amounts to keep an average valence state of +2, equal to that of Sn. X-ray photoemission spectroscopy measurements have confirmed the presence of this mixed-valence state in the closely related material In-doped PbTe~\cite{Drabkin_Charge_1982}. The high energy of the In(2+) valence state, compared to the +1 and +3 valence states, promotes Cooper pair formation, a phenomenon known as negative-$U$ superconductivity, referring to the associated negative energy of having two electrons at the same site~\cite{Anderson_Model_1975, Shelankov_Mixed-valence_1987, Varma_Missing_1988}. The consensus in literature is that this mechanism is responsible for the increase in $T_\text{c}$ of over an order of magnitude in \insnte (with $x$ > 0.02), compared to SnTe~\cite{Erickson_Enhanced_2009, Erickson_Anomalous_2010, Haldolaarachchige_Anomalous_2016, Kobayashi_Enhanced_2018, Denisov_High-field_2025}. Interestingly, this mechanism is confirmed to be responsible for the onset of superconductivity in the closely related material PbTe, when it is doped with thallium, another valence-skipping element~\cite{Dzero_Superconductivity_2005, Matsushita_Evidence_2005, Matsushita_Type-II_2006, Nakayama_Doping_2008}.

From the previous discussion, it seems likely that we have serendipitously, but reproducibly, fabricated an \insnte film through indium diffusion from the substrate. To further investigate this possibility, and the resulting indium concentration gradient, cross-sectional transmission electron microscopy combined with energy-dispersive X-ray spectroscopy (EDX) was performed on both devices (Figure~\ref{fig:TEM-EDX}). Notably, both the Hall bar and loop devices show good crystalline quality of the nanowires (Figure~\ref{fig:TEM-EDX} a,b,e,f). Additionally, at the SnTe-InP interface, small-scale roughness is observed, which may facilitate local intermixing mediated by indium dangling bonds at the substrate interface. Such interfacial effects could enable the formation of a thin In-doped SnTe region, providing a natural explanation for the observed superconductivity.

For the Hall bar device, aluminium is detected on the top facet, originating from the AlO\textsubscript{x} capping layer. Importantly, no elements originating from the contacts (and that are associated with introducing superconductivity) are observed within the nanowire. For both devices, however, no clear evidence of indium diffusion into the SnTe nanowire is found above the detection limit of a few percent. The latter is relatively high due to spectral overlap between In, Sn, and Te peaks.

\begin{figure}
\includegraphics[width=\linewidth]{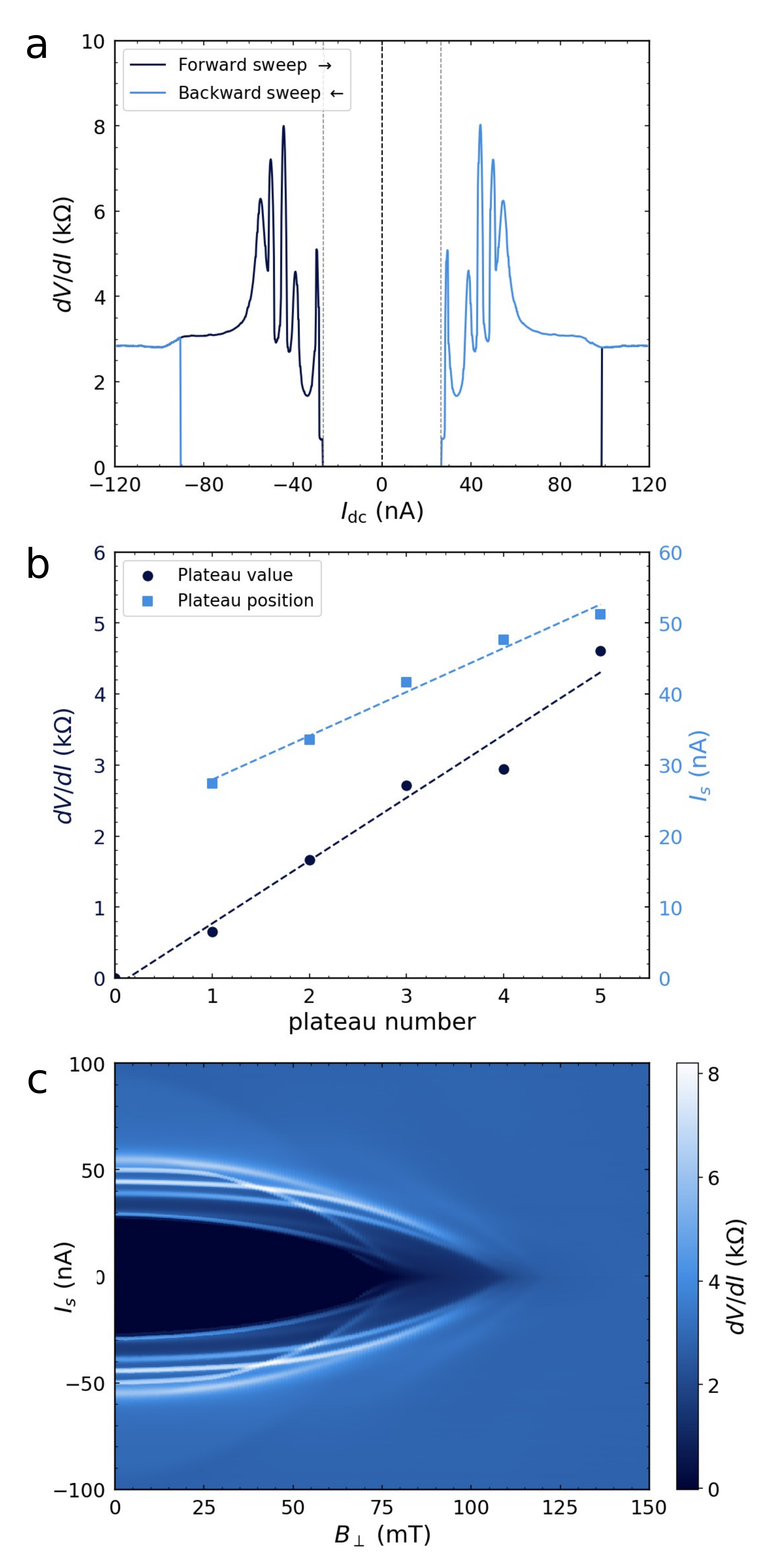}
\caption{Vortex flow channels. (a) Four-terminal longitudinal differential resistance $dV/dI$ against dc-current $I_\text{dc}$ for Hall-bar device 2. Measurement taken at base-temperature (8~mK) without magnetic field. The red and blue traces correspond to positive and negative sweep directions respectively. (b) The plateau resistance values in between the observed resonances of Figure a (black) and the positions of the plateaus (red). The dashed lines represent linear fittings. (c) Two-dimensional mapping of the resistance in (a) against out-of-plane magnetic field $B_\text{z}$. Measurement taken at $T$=50~mK.}
\label{fig:Figure3}
\end{figure}

Figure~\ref{fig:Figure3}a reveals multiple resonant features in the differential resistance of a Hall bar device as a function of source–drain current $I_\text{sd}$. These resonances appear in the intermediate regime between the zero-resistance state and the normal resistive state and are linearly spaced as a function of DC current in the Hall bar (see Figure~\ref{fig:Figure3}b and Supplementary~\ref{Suppl:Resonances}). This behaviour contrasts with the characteristic inverse voltage dependence expected for multiple Andreev reflections in Josephson junctions~\cite{Octavio_Subharmonic_1983}. Additionally, a scenario involving switching of distinct \insnte domains is unlikely, as such switching events would not naturally lead to a regularly spaced sequence of resonances. We also evaluated McMillan–Rowell and Tomasch oscillations as possible origins~\cite{Tomasch_Geometrical_1965,Rowell_Electron_1966}, but the magnitude of the resonance spacing could not be reconciled with either mechanism.

Therefore, we rather attribute the resonances to vortex dynamics within the superconducting channel. As $I_\text{sd}$ increases, vortices that are initially pinned become de-pinned, forming vortex channels that contribute discrete increments of resistance. Each resonance thus corresponds to the formation of an additional vortex channel, progressively increasing the resistance until the normal state is reached~\cite{Mcnaughton_Causes_2022}. The resistance values corresponding to the plateaus between adjacent resonances are plotted in Figure~\ref{fig:Figure3}b, showing a linear dependence (except plateau 4), consistent with the sequential opening of vortex channels. Moreover, this behaviour supports the interpretation of type-II superconductivity in \insnte as the origin of the observed phenomena.

Upon application of an out-of-plane magnetic field, the current values of all transitions  gradually diminish (Figure~\ref{fig:Figure3}c), until an upper critical field, $H_{c2}$, is reached between 100 to 150~mT. One of the resonances (plateau number 4), which also deviates from the linear trend in Figure~\ref{fig:Figure3}b, exhibits an anomalous evolution and crosses the other resonances at larger magnetic fields. Because this feature was only observed in a single device, additional experiments will be required to clarify its origin.

\begin{figure*}[t]
    \centering
    \includegraphics[width=\linewidth]{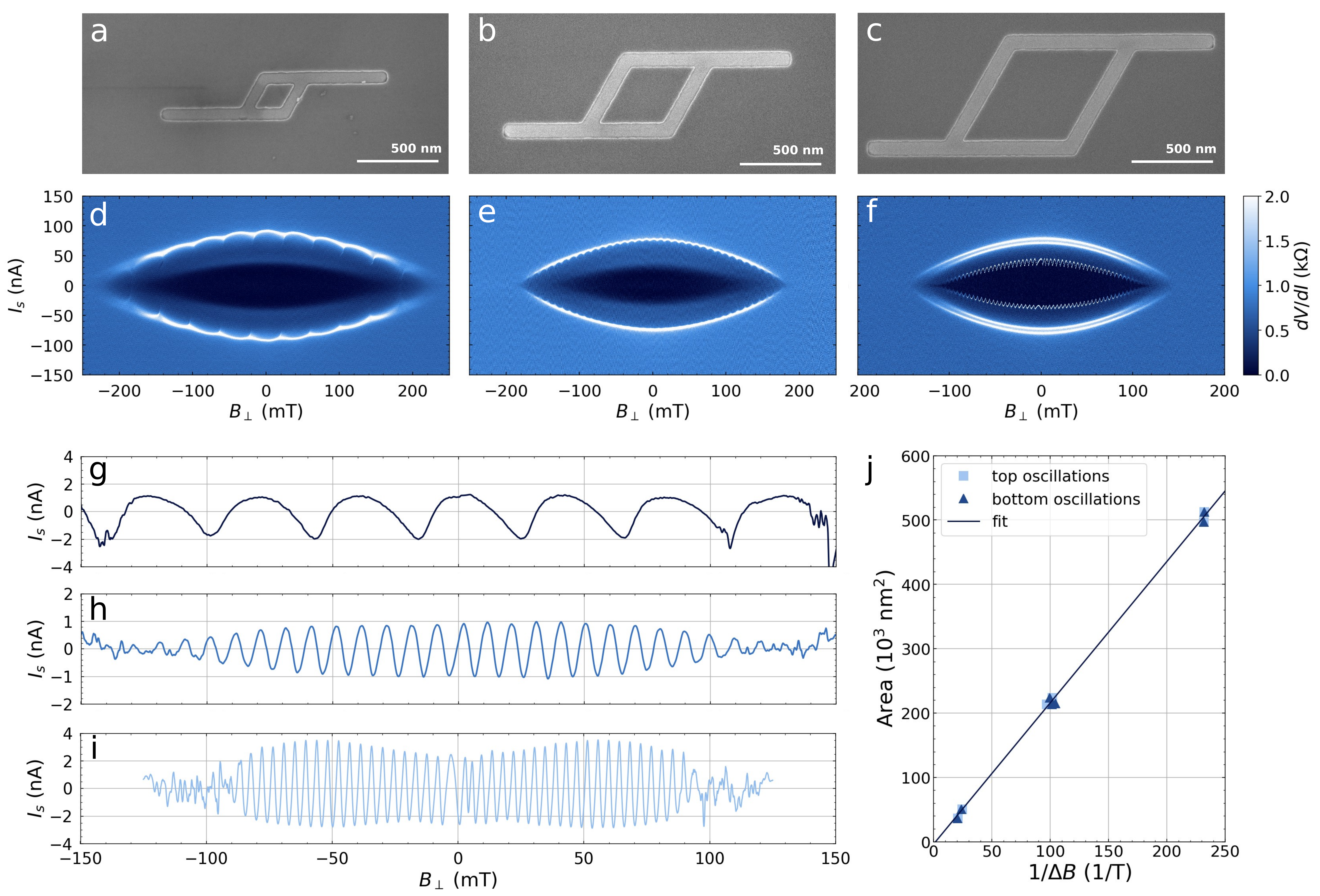}
    \caption{\textbf{Little-Parks oscillations in SnTe nanowire loops}. (a-c) SEM images of three measured devices, representative of the three investigated loop sizes. (d-e) The corresponding measured differential resistance $dV/dI$ in color scale versus magnetic field  $B_{\perp}$  and $I_\text{S}$. The oscillation frequency increases with the loop area. (g-i) Isolated oscillations in critical current against the applied magnetic field, with a subtracted background. (j) Measured oscillation frequencies versus measured effective loop area for all 8 measured devices. The dashed line has a slope of h/e (normal flux quantum) while the solid line has a slope of h/2e (superconducting flux quantum). The data matching h/2e thus points towards Little-Parks oscillations. Dashed line  corresponds to a slope of h/e.}
    \label{fig:LP}
\end{figure*}

Knowing that SnTe is supposed to be a topological crystalline insulator, the natural question arises whether the superconductivity in the system is of topological nature. As we have discussed, indium doping at the bottom surface of the SnTe film is the likely culprit for the induced superconductivity in the system. Indeed, the EDX analysis does not show In throughout SnTe channel, and we infer that the top surface of the nanowire is not intrinsically superconducting, but proximitized by the bottom part of the wire. Consequently, this material system may be an interesting candidate for topological superconductivity. We therefore study the current-phase relation in the loop-devices to verify if signatures of unconventional order-parameter symmetries can be observed.

A striking feature in the magnetoresistance, is the oscillation in the critical current $I_\text{c}$ observed for loop devices. In Figure~\ref{fig:LP}a-c top-view SEM images of three selective-area grown loops of different sizes are shown, before the deposition of any contact material. The period of the oscillations in $I_\text{c}$ observed in these devices (as shown in Figure~\ref{fig:LP}d-f) scale inversely with the loop area, consistent with the Little--Parks effect in superconducting loops. In this case, the quantization condition arises from the single-valuedness of the phase of the macroscopic superconducting ground state upon traversing around the loop. In an external magnetic field, this results in fluxoid quantization. The fluxoid is defined as the sum of magnetic fluxes through the loop originating from the external magnetic field and the resulting Meissner field. This fluxoid quantization manifests itself as a field-periodic oscillation in the critical current. For a detailed derivation, see Supplementary~\ref{Suppl:LP-derivation}.

To study the Little-Parks oscillations in our data in detail, we isolate the critical current as a function of the magnetic field from the full magnetoresistance maps, and subtract the background envelope current (see Figure~\ref{fig:LP}g-i). Using a Fourier transform, the oscillation frequency in magnetic field ($1/\Delta B$) is determined.
The area inside the loops is measured from the SEM images shown in figure  of the loops measured from~\ref{fig:LP}a-c. However, the effective area penetrated by the magnetic flux is slightly larger than the physical area due to flux focussing~\cite{Ketchen_DC_1984}. A simple model for this effective area $A_{\textrm{eff}}$ has proven to be successful for SQUID devices with varying geometries \cite{Brojeny_Magnetic-field_2003}:
\begin{equation}
    A_{\textrm{eff}}=CA_{h}\sqrt{\cfrac{A_{l}}{A_{h}}},
    \label{eq:Aeff}
\end{equation}
where $A_l$ is the outside area of the loop and $A_h$ the area of the center hole; $C$ is a constant on the order of unity. Plotting the obtained effective area for all measured loop devices against the measured $I_\text{c}$ oscillation frequency in magnetic field, yields the linear trend shown in Figure~\ref{fig:LP}j. This slope is equal to the flux quantized in the loop, multiplied by C: $A_{\textrm{eff}} = C\cdot\Phi_0/\Delta B $. Fitting this model to our data of seven loop devices, as is shown in  Figure~\ref{fig:LP}j, yields C = 1.06 and $\Phi_0=\frac{h}{2e}$, the superconducting flux quantum. This supports our hypothesis that the observed oscillations are Little-Parks oscillations.

For superconducting loops with sign-changing order parameters (such as the \textit{p}-wave symmetry, often associated with topology), half flux quanta should be present, leading to a shift of $\pi$ in Little-Parks oscillations. A minimum in $|I_c|$ should occur at zero field, instead of a maximum~\cite{Almoalem_Evidence_2022}. From the symmetry of our data in positive and negative magnetic field directions, it is concluded that the self-inductance of the devices is small enough to accurately determine the zero-field value, except for the largest loop size. For the oscillations observed in the smaller loops we detect a maximum of the critical current at zero field which is not consistent with topology-induced $\pi$-shifts. It is, therefore, likely that the observed superconductivity is dominantly carried by the trivial bulk states, thus obscuring any topological features.

To conclude, we observe superconductivity in selective-area grown SnTe nanowire devices on an InP substrate. Consistently, a flux-flow regime is observed before the wire turns normal for increased currents,  indicating  type-II superconductivity. This also explains  multiple resistance plateaus at increasing bias currents  in some of the devices. The latter are attributed to vortex flow effects, which typically occur only in type-II superconductors. Oscillations in the critical current as a function of magnetic field in nanowire loop devices are explained by the Little-Parks effect.

From TEM/EDX data and a  process of elimination, we identify indium diffusion from the InP substrate into the SnTe nanowire as the most plausible origin of the superconductivity. While no direct evidence of indium incorporation appears  above the detection limit of the EDX measurements, the interfacial roughness at the SnTe-InP interface might facilitate local intermixing and the formation of a thin In-doped SnTe region near the substrate. Such a region would provide a natural explanation for the intrinsic superconductivity observed in these devices.

Our material system might become a platform for topological superconductivity, as the part of the wire close to the substrate is believed to be intrinsically superconducting, whereas the top part may still be in the TCI phase. However, a $\pi$-shift in the Little-Parks oscillations associated with non-trivial superconductivity has not been observed in any of the loop devices. It is likely that the trivial bulk \insnte dominates the superconducting transport, obscuring any signals associated with the non-trivial topology. These observations outline the path forward; once the bulk conduction is suppressed, the in-situ indium diffusion should be repeated and the superconductivity could retain its topological nature.

\section*{Methods}
\subsection*{SnTe SAG nanostructure growth}
An InP(111)A substrate was covered with a SiN$_x$ mask, in which openings were etched down to the substrate. Under appropriate growth conditions, SnTe nucleated selectively within the openings and grew epitaxially, resulting in monocrystalline nanostructures with a (111) top surface. Nanowires with widths as small as 50 nm and lengths exceeding 10~\textmu m were grown with high yield. This selective-area growth method enabled the formation of complex nanowire networks and devices, which was utilised to fabricate the investigated Hall bars and loops.

\subsection*{FIB/TEM/EDX}
Cross-sectional TEM samples have been made using a Thermo Fisher Helios 5 Dual Beam Focused Ion Beam (FIB) system, using a lift-out sample preparation scheme. Electron Beam Induced Deposition (EBID) of C and Ion Beam Induced Deposition (IBID) of Pt/C have been applied to protect the location of interest from preparation damage. Subsequent TEM studies have been performed using an abberation corrected JEOL JEM200F, operated at 200 kV, equipped with a 100 mm\textsuperscript{2} SDD Energy Dispersive X-ray Spectroscopy (EDS) detector. Imaging was performed in scanning TEM mode, simultaneously using a High Angle Annular Dark Field (HAADF) and a Bright Field (BF) detector.

\subsection*{Device fabrication}
\label{subsec:fabrication}
The Hall bar devices and the contact loop were fabricated in different laboratories, resulting in slight differences in the fabrication procedures.

For the Hall bar devices, the samples were first capped \textit{in situ} by electron-beam evaporation with a 2 nm AlO\textsubscript{x} layer. Electron-beam lithography (EBL) was then performed using a single layer of PMMA A6 resist. After exposure, the resist was developed at low temperature in MIBK, followed by a rinse in IPA. Prior to metallization, the exposed surface was cleaned by a short O\textsubscript{2} plasma treatment for 15 s and a low-power Ar sputter etch for a total duration of 480 s. Subsequently, a 10 nm Ti / 100 nm Au contact stack was deposited \textit{in situ} by electron-beam evaporation. The remaining resist was removed by cold lift-off in acetone.

For the loop devices, EBL was carried out using PMMA A4 resist spin-coated at 4000 rpm and baked for 3 min at 160~$^\circ$C. After exposure, the resist was developed at low temperature in MIBK for 60 s, followed by a 30 s rinse in IPA. The exposed surface was cleaned by a low-power Ar sputter etch for 40 s. A two-step metal deposition was then employed, consisting of sputter deposition of 1 nm Ti and 4 nm Pd, followed by electron-beam evaporation of an additional 25 nm Pd. Finally, lift-off was performed in DMSO at 50~$^\circ$C. This two-step deposition process was used to minimize sidewall deposition.

\subsection*{Measurements}
The Hall bar devices were measured in a dilution refrigerator with a base temperature of 8~mK using a Zurich Instruments MFLI lock-in amplifier in combination with a Physics Basel SP1'004 low-noise differential amplifier. During magnetic-field sweeps, the sample temperature was stabilised at 50~mK.\\
The loop devices were measured in a separate dilution refrigerator with a base temperature of 40~mK. A Moku:pro DC lock-in amplifier (averaging over 1024 IV traces) was used in combination with a TUDelft QT-designed IVVI-Dac2 measurement rack.

\section*{Acknowledgements}
The authors would like to thank Sander Schellingerhout, Valentyn Volobuev, Sebastian Miles and Denny Lamon for useful discussions. Thanks are owed to Daan Wielens, Frank Roesthuis, Mario Cignoni and Dennis van der Bovenkamp for practical assistance in the labs and to Klaas Koopmans for his assistance in the device design.\\
This work was financially supported by the Dutch Organization for Scientific Research through the project “HOTNANO” (OCENW.GROOT.2019.004), the research program "Materials for the Quantum Age (QuMat)" (Registration No. 024.005.006), and Solliance and the Dutch province of Noord-Brabant for funding of the TEM facility.   

\newpage
\bibliographystyle{ieeetr}
\bibliography{references} 
\newpage
\appendix
\onecolumn
\section*{Supplementary}

\section{Measurement over metallic lead}
\label{Suppl:Contacts}
\begin{figure}[h]
    \centering
    \includegraphics[width=0.9\linewidth]{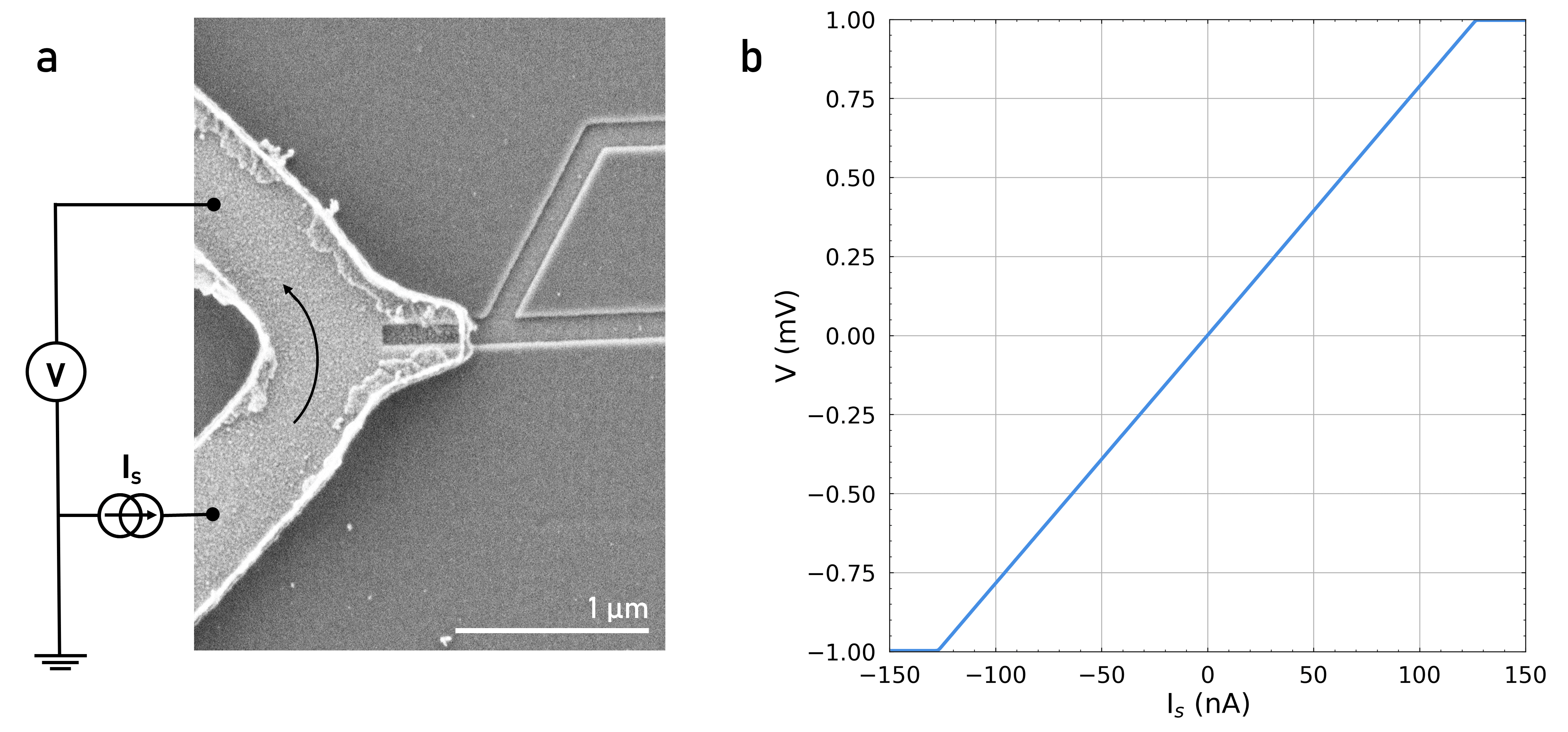}
    \caption{Two-terminal IV curve through one of the contacts on a loop device. \textbf{(a)} SEM image of the relevant part of the device, with the measurement setup indicated schematically. \textbf{(b)} Plot of the measured IV curve. The linear (ohmic) curve does not exhibit a plateau for low currents, indicating that the titanium/palladium metal stack used for the contacts is not intrinsically superconductive.}
    \label{Suppl_Fig:Contact resistance}
\end{figure}

The loop devices in this study were contacted in a quasi-four-point configuration: the nanowire was contacted on one point on each side, but is split in two directly next the point of contact (see Figure \ref{Suppl_Fig:Contact resistance}a). Measuring in this configuration excludes the line resistance, but not the contact resistance due to the metal-nanowire interface.\\
Here, this geometry is exploited to measure the resistance of the leads in a two-point dc measurement. The resulting IV curve is shown in Figure \ref{Suppl_Fig:Contact resistance}b; it is clearly Ohmic, as expected for a strip of metal. This measurement demonstrates that the titanium/palladium leads used for these devices are not intrinsically superconducting, excluding this as a cause of the superconducting behaviour.

\section{Hall measurement}
\label{Suppl:Hall}

\begin{figure}[H]
    \centering
    \includegraphics[width=0.6\linewidth]{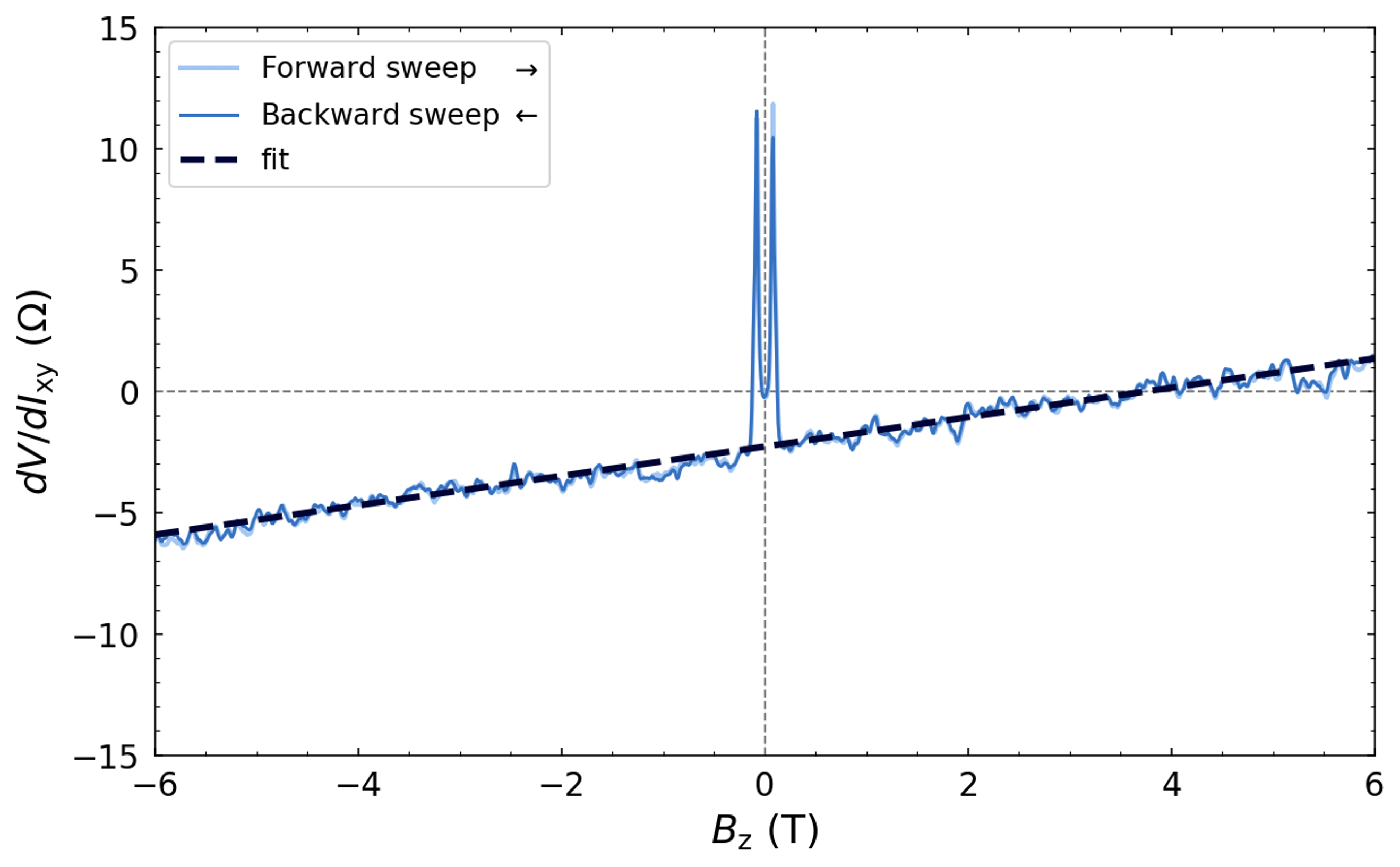}
    \caption{Four-terminal tranverse differential resistance as function of out-of-plane magnetic field for Hall-bar device 2, measured at T=50 mK. The red and blue traces correspond to opposite magnetic-field sweep directions. The green line indicates the linear fit used to extract the carrier density.}
    \label{Suppl_Fig:Hall measurement}
\end{figure}

Hall measurements were performed on both Hall bar devices discussed in the main text to determine the carrier density. An example of such a measurement for device 2 is shown in Figure \ref{Suppl_Fig:Hall measurement}. Above the critical field, the Hall resistance exhibits a linear dependence on the magnetic field, from which the carrier density was extracted using a linear fit. The resulting carrier densities are $2.31\cdot10^{20}$ cm\textsuperscript{-3} and $2.05\cdot10^{20}$ cm\textsuperscript{-3} for devices 1 and 2, respectively. These relatively high carrier densities are consistent with previous reports \cite{Volobuev_Giant_2017} and can be attributed to the presence of Sn vacancies in the nanowires.

In the low-field regime, below the critical field, the Hall signal deviates from the linear behavior due to the onset of superconductivity. In addition, small-scale fluctuations are observed as a function of magnetic field. Because these fluctuations are reproducible for both upward and downward field sweeps, they are attributed to mesoscopic conductance fluctuations.
Previous studies have shown that superconductivity in Sn\textsubscript{1-$\delta$}Te can emerge when the Sn-vacancy concentration is sufficiently high \cite{Hulm_Superconducting_1968,Allen_Carrier_1969}. For the $T_\text{c}$ of 100–200 mK observed in our Hall-bar devices, this would correspond to a required carrier density on the order of $1$-$2\cdot10^{21}$ cm\textsuperscript{-3}. Our measured Hall densities of approximately $2\cdot10^{20}$ cm\textsuperscript{-3} are therefore an order of magnitude too low, suggesting that the observed superconductivity does not originate from Sn\textsubscript{1-$\delta$}Te. Moreover, the established trend of increasing $T_\text{c}$ with carrier density in Sn\textsubscript{1-$\delta$}Te does not apply to our devices.

\section{Evaluation of the superconducting coherence length}
\label{Suppl:Xi}

Assuming a carrier density of $p = 2 \cdot 10^{20}\text{ cm}^{-3}$, taking into account the fourfold valley degeneracy at the L-points in the Brillouin zone ($N_\text{v}=4$) and assuming $m^*\approx 0.15m_0$ (with $m_0$ the electron's rest mass)~\cite{chen2013importance} we estimate the Fermi k-vector and Fermi velocity:
\begin{equation}
\begin{split}
k_\text{F} = \left( 3\pi^2 \frac{p}{N_\text{v}} \right)^{1/3} \approx 1.1 \times 10^9\text{ m}^{-1} \\
v_\text{F} = \hbar k_\text{F} / m^* \approx 8.8 \times 10^5\text{ m/s}
\end{split}
\end{equation}
We use the BCS gap $\Delta = 1.764 k_\mathrm{B}T_\mathrm{C}$ with the lower limit of observed critical temperatures $T_\mathrm{C}\approx 200~\text{mK}$ to calculate the superconducting coherence length in the clean limit: 
\begin{equation}
\xi_0 = \frac{\hbar v_\text{F}}{\pi \Delta} \approx 6.1\ \mu\text{m}.
\end{equation}
However, the high carrier density due to the Sn vacancies implies a strong suppression of ballistic transport. To calculate the elastic mean free path $l$, we first determine the resistivity $\rho$ for the device shown in Fig.~\ref{fig:SEM+SC}a and consider this as as our general material estimate. Using the normal state resistance $R_\text{N}\approx3.0~\text{k}\Omega$, nanowire width $w=120~\text{nm}$, nanowire height $h=60~\text{nm}$, and nanowire length $L\approx 5.0~\mu\text{m}$, we find $\rho = R_\text{N} \cdot w \cdot h / L \approx 4.0 \cdot 10^{-6}\ \Omega\cdot\text{m}$. Plugging this into the Drude transport model we find 

\begin{equation}
l = \frac{m^* v_\text{F}}{p e^2 \rho} \approx 6.0 \text{ nm}\text{,}
\end{equation}
clearly indicating that $l \ll \xi_0$ and we reside deep within the diffusive dirty-limit.

Given that we are in the low temperature limit $T\rightarrow 0$, the coherence length in this regime converges to the geometric mean of the clean limit and the elastic mean free path~\cite{Tinkham}:
\begin{equation}
    \xi_\text{D}\approx 0.85\sqrt{\xi_0l}\approx160~\text{nm}.
\end{equation}

The limit $\xi_\text{D} \ll L$ implies a strongly decaying superconducting wave function: $\Psi(x) \propto \exp(-x/\xi)$, with $x$ indicating the position along the nanowire between $0$ and $L$. Indeed, at the middle of the nanowire ($x=2.5~\mu\text{m}$) a suppression factor of $\exp(-2.5\mu\text{m} / 160\text{ nm}) \approx 1.8 \times 10^{-7}$ applies, confirming that superconducting transport does not take place by means of a Josephson current.

\section{Absence of Josephson behavior in experiments}
\label{Suppl:Shapiro}
Let us take the opposite point of view and try to prove that the superconductivity is not intrinsic to nanowires, then the observed superconductivity can only be induced by the contacts. The observation of Little-Parks oscillations indicates that the superconducting coherence is preserved throughout the entire loop. This can only be the case if the regions of the nanowire proximitized by supposedly superconducting contacts overlap, allowing coherent transport between the contacts: a Josephson junction \cite{Tinkham, Josephson_Possible_1962, Josephson_Supercurrents_1965}.

\begin{figure}[h]
    \centering
    \includegraphics[width=0.9\linewidth]{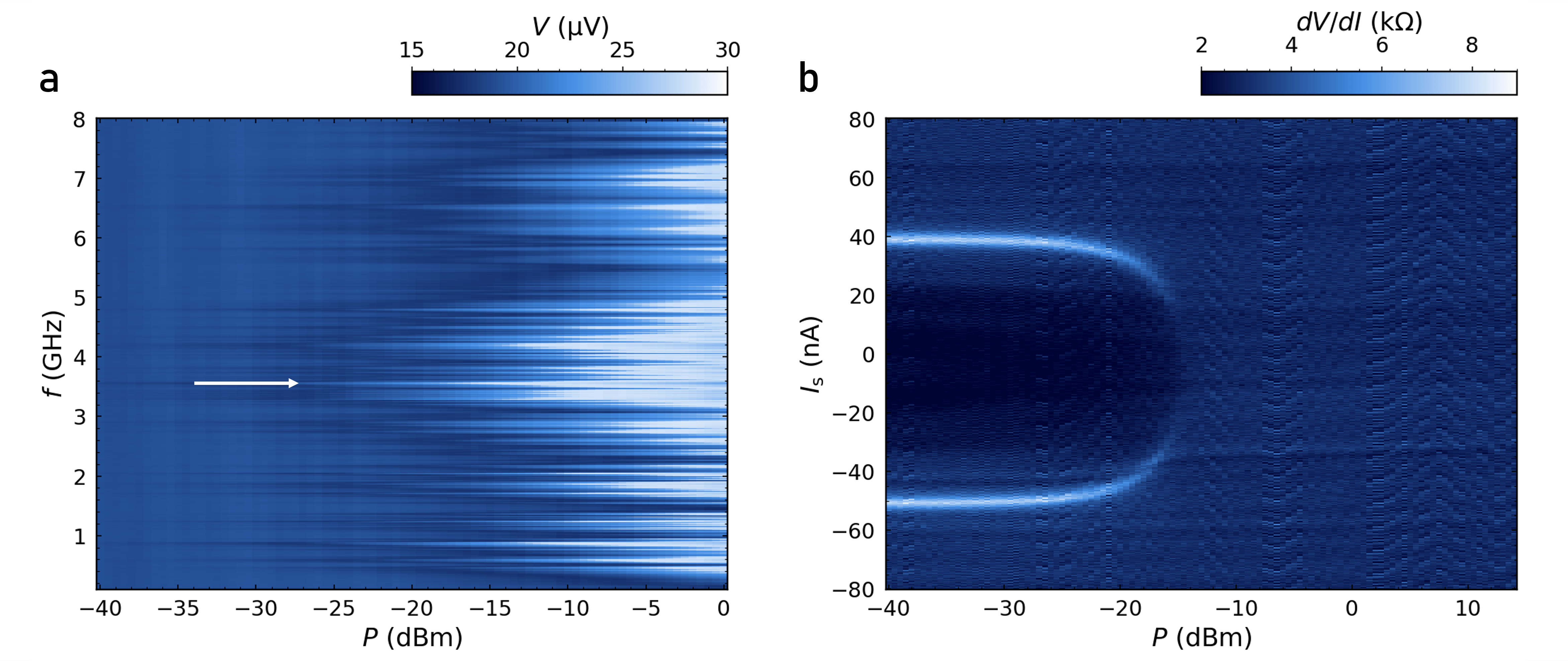}
    \caption{Shapiro experiment on one of the loop devices. \textbf{(a)} The generated voltage over the device as a function of the applied rf frequency and power. The absorption of radiation is significant over the entire spectrum. \textbf{(b)} The applied current vs radiation power map at the rf frequency indicated with a white arrow in figure \textbf{(a)}. The superconducting gap closes due to heating effects from the absorbed radiation; no Shapiro steps were observed.}
    \label{Suppl_Fig:Shapiro}
\end{figure}

A universal feature of Josephson junctions is the spontaneous generation of a quantized voltage over the junction, as a result of the absorption of microwave photons: Shapiro steps \cite{Shapiro_Microwave_1967}. We have mounted an rf antenna right over the chip with nanowire devices (distance approx. 1 cm). Figure \ref{Suppl_Fig:Shapiro}a shows the generated voltage over the junction as a function of applied power and rf frequency. Absorption of radiation is significant over the entire frequency range. A frequency with particularly strong coupling to the sample is selected (in this case 4.4 GHz, indicated with a white arrow), to make the Shapiro map: steps in voltage (or equivalently, differential resistance) should occur when a current and rf radiation are applied.\\
As can be seen in Figure \ref{Suppl_Fig:Shapiro}b, these steps do not occur. The superconducting gap closes due to heating effects at high applied rf power, but that is the only discernable in the data. As the presence of Shapiro steps is a general feature of Josephson junctions, this therefore supports the claim that the superconductivity in our devices is \textit{not} due to the contacts. This experiment was repeated for multiple radiation frequencies and repeated on two other devices, all with similar results.

\section{EDX elemental maps}
\label{suppl:EDX}
\begin{figure}[H]
\centering
\includegraphics[width=0.85\linewidth]{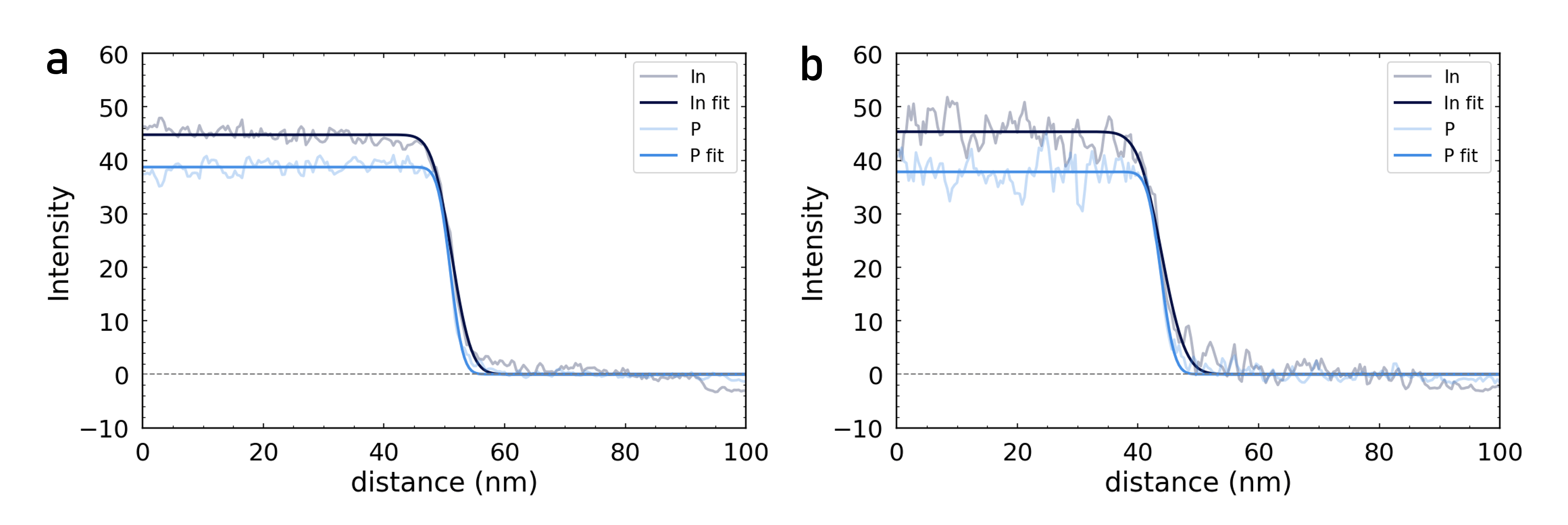}
\caption{Line traces of the In-L and P-K EDX intensities across the substrate-nanowire interface for the Hall-bar device (\textbf{a}) and AB-loop device (\textbf{b}) discussed in the main text. The solid lines show the corresponding error-function fits.}
\label{Suppl_Fig:EDX line traces In and P}
\end{figure}

Figure \ref{Suppl_Fig:EDX line traces In and P} shows line traces of the In-L and phosphorus-K peaks extracted from the EDX spectra presented in the main text (indicated by dashed lines). To estimate the possible broadening of the In signal at the substrate–nanowire interface with respect to the phosphorus signal, the traces were fitted using an error-function profile, used to describe a step that is broadened by instrumental resolution or inter-diffusion:

\begin{equation}
    I(x)=c_{\text{s}}+\frac{c_{\text{n}}-c_{\text{s}}}{2} \left[ 1+\text{erf}\left( \frac{x-w}{\sqrt{2}\sigma} \right) \right],
    \label{eq:erf-function}
\end{equation}
where $c_{\text{s}}$ and $c_{\text{n}}$ represent the intensity levels of the substrate and nanowire regions, respectively, $w$ the position of the interface, and $\sigma$ the broadening parameter that describes the transition width. During the fitting procedure, the baseline inside the nanowire region was fixed to zero for both traces to allow a direct comparison of the interface profiles. 

Comparison of the resulting fits for In and P suggests that, for both the loop and Hall bar device, the In signal may extend slightly further into the nanowire region than the P signal. However, a reliable quantitative comparison is hindered by the fact that the In signal also exhibits a higher background level inside the substrate region. Furthermore, the strong spectral overlap between the In-L and Sn-L peaks can lead to artificial In counts within the SnTe nanowire region, which may artificially broaden the step in the EDX line trace. For these reasons, we refrain from drawing firm conclusions regarding measurable In diffusion on the basis of these spectra. Nevertheless, the absence of a resolvable In signal throughout the nanowire excludes a large-scale In diffusion into the entire SnTe nanowire.

\begin{figure}[H]
    \centering
    \includegraphics[width=0.9\linewidth]{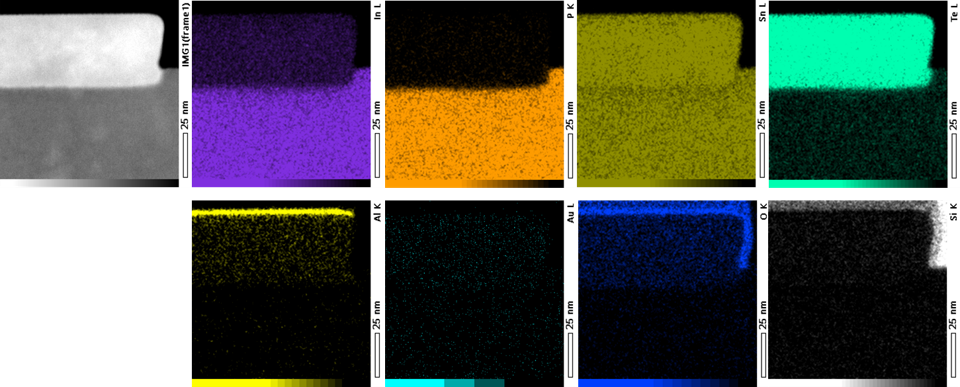}
    \caption{Energy-dispersive X-ray spectroscopy (EDX) spectra of the detected elements for the Hall-bar device discussed in the main text.}
    \label{Suppl_Fig:EDX elements Hall-bar}
\end{figure}

\begin{figure}[H]
    \centering
    \includegraphics[width=0.9\linewidth]{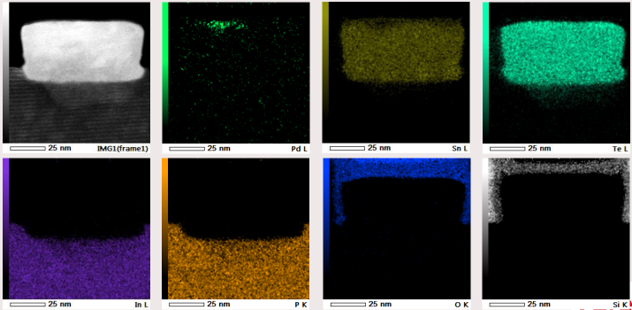}
    \caption{Energy-dispersive X-ray spectroscopy (EDX) spectra of the detected elements for the AB-loop device discussed in the main text.}
    \label{Suppl_Fig:EDX elements AB-loop}
\end{figure}

\begin{figure}[H]
    \centering
    \includegraphics[width=0.9\linewidth]{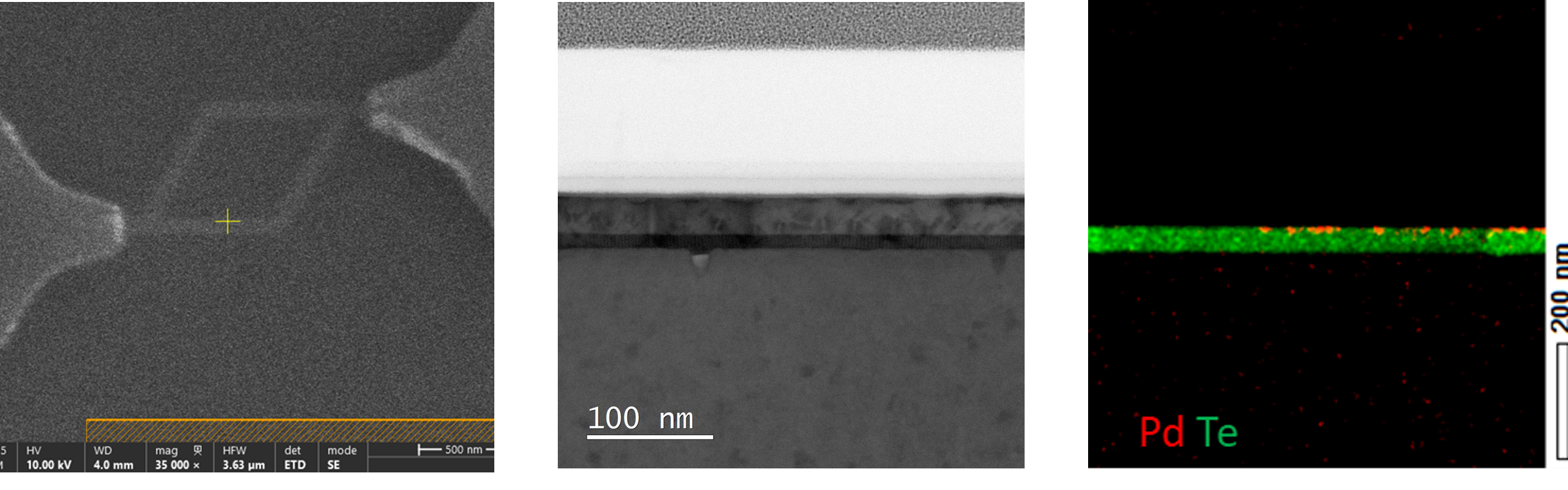}
    \caption{TEM image and corresponding EDX spectra of a longitudinal cross-section along the axis of a SnTe nanowire from an AB-loop device. Palladium islands are observed on top of the nanowire surface.}
    \label{Suppl_Fig:Longitudinal TEM-EDX}
\end{figure}

Figures \ref{Suppl_Fig:EDX elements Hall-bar} and \ref{Suppl_Fig:EDX elements AB-loop} show the individual EDX spectra of the detected elements for both devices discussed in the main text. The spectra confirm the presence of the expected elements originating from the substrate, nanowire, and contact or capping layers. In particular, peaks corresponding to Sn and Te are observed in the nanowire region, while In and P originate from the InP substrate. For the Hall-bar device, the detected Al corresponds to the capping layer of AlO\textsubscript{x}. However, for the AB-loop device, some Pd is also detected on top of the wire away from the actual Ti/Pd contacts. To investigate the latter in more detail, Figure \ref{Suppl_Fig:Longitudinal TEM-EDX} shows TEM and EDX data obtained from a longitudinal cross-section along the nanowire axis of an AB-loop device that exhibited similar superconducting behavior. In the EDX mapping, Pd islands are clearly observed on top of the SnTe nanowire. Since these islands form a discontinuous layer and are only present on the top surface of the nanowire, they are unlikely to originate from Pd-diffusion from the Ti/Pd contacts (which could otherwise lead to the formation of superconducting PdTe\textsubscript{2}, as reported for other topological insulators based on Te \cite{Kononov_One_2020,Bai_Novel_2020}). Instead, the Pd islands are most likely the result of fabrication or sample-preparation processes, such as lift-off or the focused ion beam (FIB) cutting procedure.

In any case, the discontinuous nature of the Pd layer strongly argues against it being responsible for the observed superconductivity. More importantly, no Pd was detected in any of the Hall bar devices, whose contacts do not contain Pd, while these devices exhibit superconducting behavior comparable to that observed in the loop devices. This further excludes a significant role for Pd in the origin of the superconductivity.

\section{Analysis of resonances}
\label{Suppl:Resonances}
\begin{figure}[H]
\centering
\includegraphics[trim=50pt 30pt 30pt 30pt, clip, width=\linewidth]{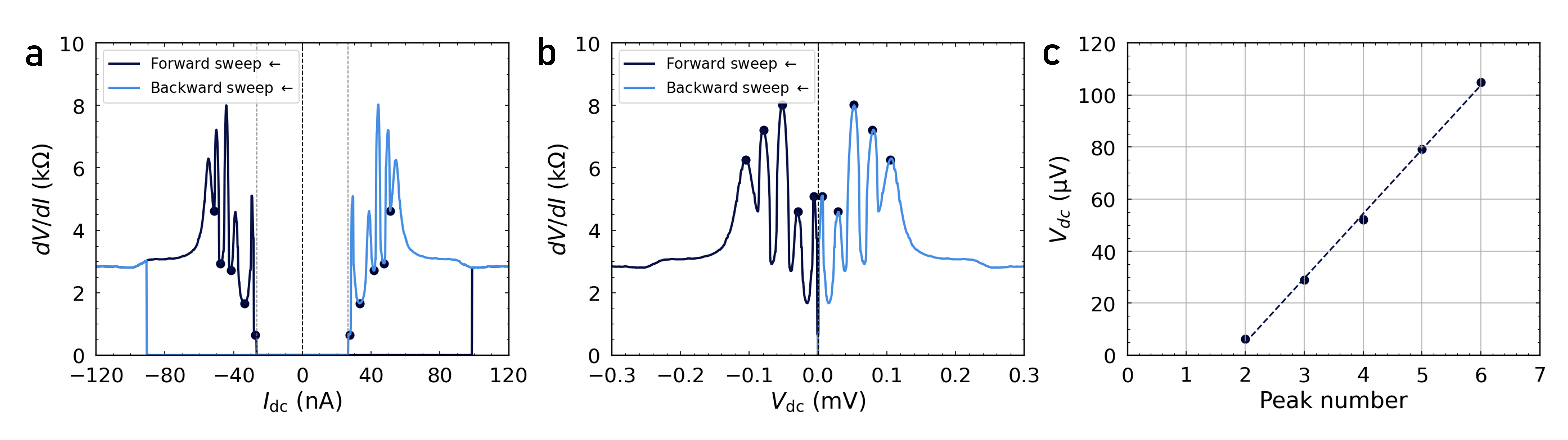}
\caption{\textbf{(a)} Four-terminal longitudinal differential resistance against dc-current for Hall-bar device 2. Measurement taken at base-temperature (8 mK) without magnetic field. The red and blue traces correspond to a forward and negative sweep direction respectively. The bright red circles indicate the local minima. \textbf{(b)} The four-terminal differential resistance of Figure \textbf{(a)} against the (integrated) dc voltage across the device. The bright red circles indicate the local maxima. \textbf{(c)} The peak position values extracted from Figure \textbf{(b)}. The dashed line represents a linear fit}
\label{fig:SI_Figure_resonances}
\end{figure}

To determine the resistance values of the plateaus and their positions, the values of $R$ and $I_\text{dc}$ are extracted from the local minima as shown in Figure \ref{fig:SI_Figure_resonances}a. For clarity, our analysis in the main text focuses on the $B_\text{z}$>0 region, although comparable resonances are also present for $B_\text{z}$<0. To express the differential resistance as function of the dc voltage, we integrated the differential resistance according to
\begin{equation}
    V_\text{dc}(I_\text{dc})=\int_0^{I_\text{dc}}\frac{dV}{dI}(I')dI',
    \label{eq:Idc-to-Vdc}
\end{equation}
yielding the $R$-$V_\text{dc}$ graph in Figure \ref{fig:SI_Figure_resonances}b.
Performing a similar analysis on the peak positions yields the linear graph in Figure \ref{fig:SI_Figure_resonances}c. However, note that upon integration, the first plateau and peak visible in Figure \ref{fig:SI_Figure_resonances}b could not be resolved. In any case, a clear linear trend is observed in the voltage positions of the resonances, contradicting multiple Andreev reflections.

\begin{figure}[H]
\centering
\includegraphics[trim=5pt 5pt 5pt 5pt, clip, width=0.85\linewidth]{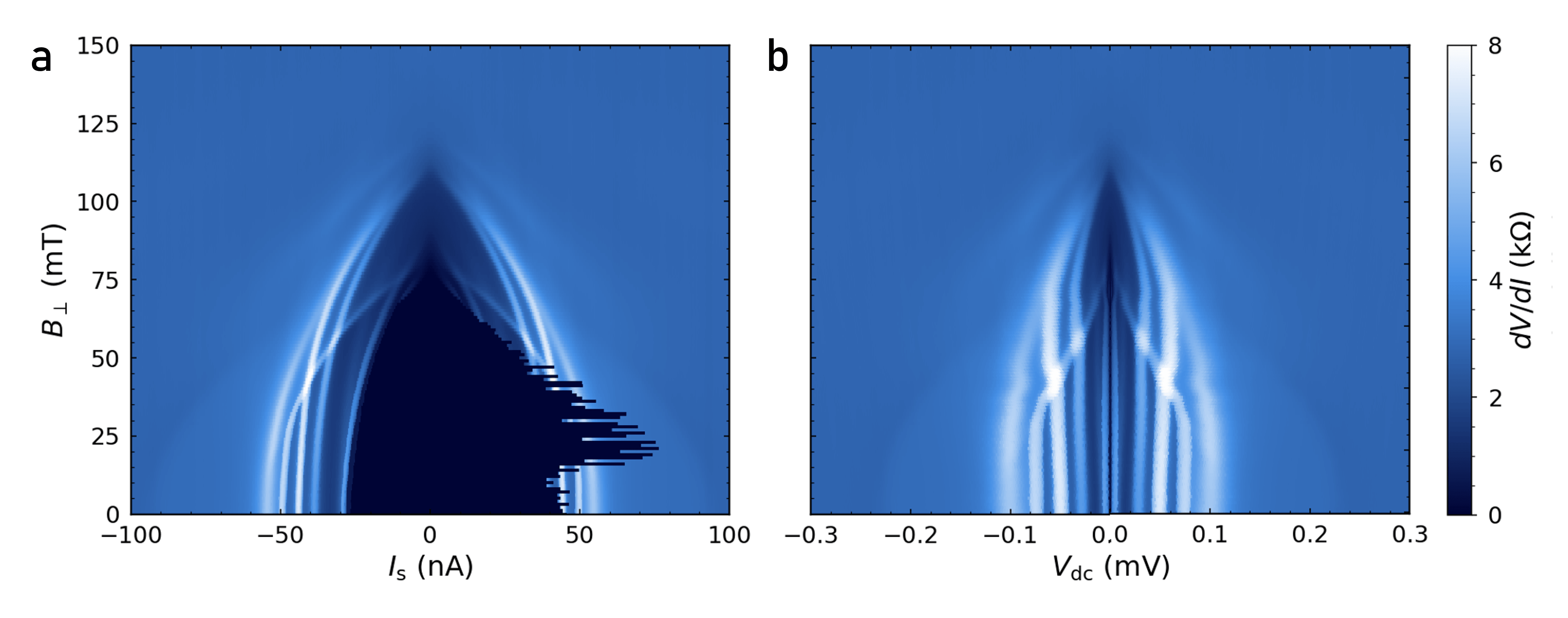}
\caption{\textbf{(a)} Two-dimensional map of the four-terminal longitudinal resistance as function of out-of-plane magnetic field for Hall-bar device 2, measured at T= 50mK. The forward sweep direction produces a clear superconducting hysteresis. \textbf{(b)} Two-dimensional map of the longitudinal resistance plotted as function of the dc voltage across the device. Hysteresis has been removed by combining the inward-sweep segments from measurements taken with opposite sweep directions.}
\label{fig:SI_Figure_resonances_magneticfield}
\end{figure}

Figure \ref{fig:SI_Figure_resonances_magneticfield}a shows the raw differential resistance data of the same Hall bar device as a function of out-of-plane magnetic field for a current sweep from negative to positive $I_\text{sd}$. In addition to the resonant features, a pronounced superconducting hysteresis is observed. To remove the effect of this hysteresis when analyzing the resonance structure, an additional measurement was performed with the sweep direction reversed. The inward sweep segments (from -$I_\text{sd}$ to zero and from +$I_\text{sd}$ to zero) of the two measurements were then combined, resulting in the hysteresis-free trace shown in Figure \ref{fig:Figure3}c of the main text. 
To express the magnetic-field map as a function of the dc voltage, we performed the same integration as given by equation \ref{eq:Idc-to-Vdc}, yielding the representation shown in Figure \ref{fig:SI_Figure_resonances_magneticfield}b. 

\section{Influence of AC excitation current on the resonance features}
\begin{figure}[H]
\centering
\includegraphics[trim=30pt 30pt 30pt 30pt, clip, width=\linewidth]{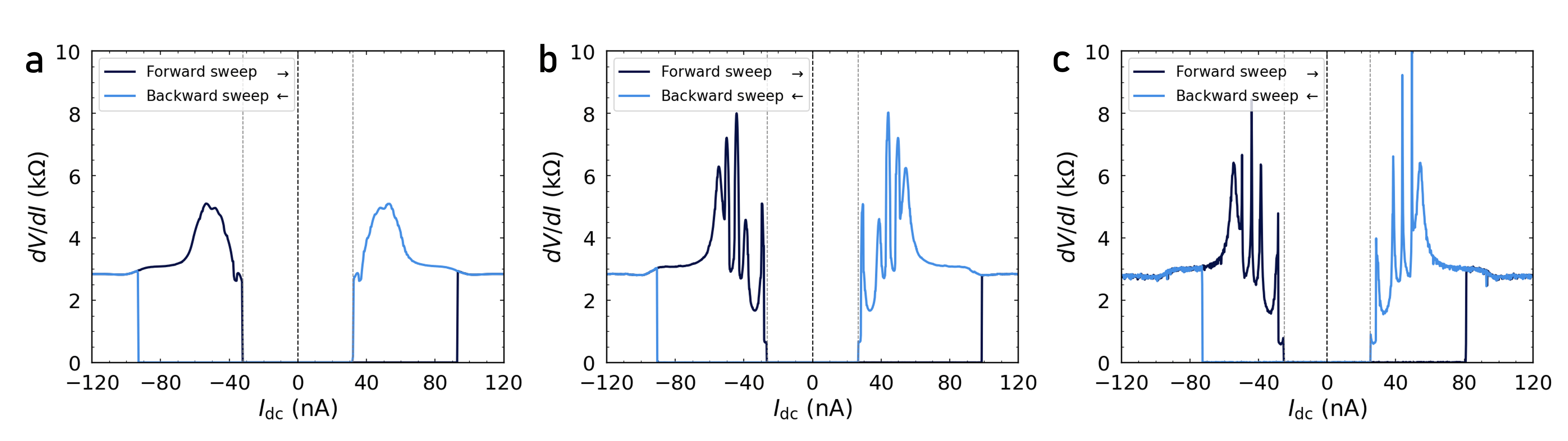}
\caption{Four-terminal longitudinal differential resistance against dc-current for Hall bar device 2, using a lock-in ac excitation current of magnitude \textbf{(a)} 5 nA, \textbf{(b)} 1 nA and \textbf{(c)} 0.1 nA.}
\label{fig:SI_Figure_resonances-vs-ac-current}
\end{figure}

Figure \ref{fig:SI_Figure_resonances-vs-ac-current} shows lock-in measurements on a Hall bar device using different rms AC excitation amplitudes (5~nA, 1~nA and 0.1~nA respectively). Decreasing the excitation current improves the resolution of the resonance features, but reduces the signal-to-noise ratio. Since the peak positions obtained with 1~nA and 0.1~nA excitation are strongly similar, an rms AC excitation current of 1~nA was used for all Hall bar measurements presented in the main text. 

\section{Little-Parks oscillations - Derivation}
\label{Suppl:LP-derivation}
A brief derivation of the Little-Parks effect is given based on the Ginzburg-Landau theory. Here we roughly follow the derivation from the seminal work by Tinkham \cite{Tinkham, Tinkham_Effect_1963}.

Flux quantization in rings or cylinders is known from normal state mesoscopic physics, of which the Aharonov-Bohm effect \cite{AharonovBohm1959Significance} is the most straightforward example.
A superficially very similar effect was observed in superconducting cylinders in the landmark experiments by Little and Parks \cite{LittleParks_Observation_1962, LittleParks_Fluxoid_1964}.  However, in superconductors it is not strictly the magnetic flux that is quantized, but the closely related \textit{fluxoid}. This fluxoid $\Phi'$ was defined in 1950 by Fritz London as the magnetic flux plus the flux induced by the screening current running over the surface of the superconductor:\\
\begin{equation}
    \Phi'= \iint_S \textbf{B}\cdot d\textbf{s} + \mu_0\oint \lambda^2 \textbf{J}_s \cdot d\textbf{l}
    \label{eq:DefFluxoid}
\end{equation}

Here we introduce the London penetration depth $\lambda = \sqrt{m^*/\mu_0n_s q^2}$ the typical depth the screening current penetrates into the bulk of the superconductor.

Introducing the vector potential $\textbf{A} = \nabla\times\textbf{B}$ and applying Stokes theorem, allows us to combine both contributions into a single line integral:
\begin{equation}
    \Phi'= \oint\left[\textbf{A} + \mu_0\lambda^2 \textbf{J}_s\right]\cdot d\textbf{l}
\end{equation}

Using the definition of the London penetration depth and expressing the supercurrent density in terms of the condensate velocity $\textbf{J}_s = n_s q \,\langle\textbf{v}_s\rangle$, this integral can be manipulated to read:
\begin{equation}
    \Phi' = \frac{1}{q}\oint\left[q\textbf{A}+m^*\textbf{v}_s  \right]\cdot d\textbf{l}
\end{equation}
In the integrant we can now recognize the canonical momentum: $\textbf{p}=q\textbf{A}+m^*\textbf{v}_s$. Quantization of momentum on loop is what will eventually lead to the quantization condition of the fluxoid. Tinkham \cite{Tinkham_Effect_1963} recognizes the radial action in this integral and therefore argues that the semiclassical Bohr-Sommerfeld quantization condition can be applied: $\oint\textbf{p}\cdot d\textbf{l} = nh$. This results in the fluxoid quantization condition:
\begin{equation}
    \begin{split}
        \Phi' & = n\frac{h}{2e} \\
              & = n\Phi_0
    \end{split}
\end{equation}
with q=2e the charge of a cooper pair. Here we define the superconducting flux quantum $\Phi_0=\frac{h}{2e}$.\\

The same result can be derived more rigorously in the framework of Ginzburg-Landau (GL) theory. GL theory describes the superconducting state in terms of an order parameter $\psi = |\psi|e^{i\varphi}$, which is closely related to the Cooper pair density $n_s=|\psi|^2$. In this approach the quantization condition arises from the single-valuedness of the order parameter: the phase $\varphi$ of the order parameter can only vary in integer multiples when traversing the loop completely:
\begin{equation}
\oint\nabla\varphi\cdot d\textbf{l}=2\pi\,n
\end{equation}
Realizing that the momentum is defined in terms of the order parameter as $\textbf{p}=\hbar\nabla\varphi$, yields:
\begin{equation}
    \begin{split}
        \Phi' & = \frac{\hbar}{q}\oint\nabla\varphi \cdot d\textbf{l} \\
              & = n\frac{2\pi\,\hbar}{2e} \\
              & = n\Phi_0
    \end{split}
\end{equation}
Which is the same quantization condition we arrived at before.

Now let's rewrite equation \ref{eq:DefFluxoid} by substituting  this quantization condition, and the previously given definitions for the London penetration depth and the supercurrent density:
\begin{equation}
    n\Phi_0 = \Phi - \oint \frac{m^*}{2e} \textbf{v}_s \cdot d\textbf{l}
\end{equation}
Integrating over the circumference of the loop (in our case 4 times the side length L), this results in the following quantization condition for the condensate velocity:
\begin{equation}
    v_s = \frac{1}{4L}\frac{h}{m^*} \left( n-\frac{\Phi}{\Phi_0} \right)
\end{equation}

Using Ginzburg-Landau theory, this can be used to express the superconducting coherence length $\xi$ in terms of the quantized flux \cite{Tinkham}:
\begin{equation}
    \frac{1}{\xi} = \frac{\pi}{2L} \left( n-\frac{\Phi}{\Phi_0} \right),
\end{equation}
which in turn is related to change in critical temperature as $\Delta T_c \propto 1/\xi^2$. The exact proportionality constant here depends on geometry and differs for the clean and dirty limits. The qualitative behaviour remains the same:
\begin{equation}
    \Delta T_c \propto \left( n-\frac{\Phi}{\Phi_0} \right)^2
\end{equation}
These periodic scallops in $T_c$ with the flux through the wire is what Little and Parks observed in their landmark experiments \cite{LittleParks_Observation_1962, LittleParks_Fluxoid_1964}.

However, in practice it is often easier to measure variations in the critical current at a constant temperature, than this change in $T_c$. The critical current density $J_c)$ is directly related to the critical temperature, according to the Landau-Ginzburg formalism \cite{Tinkham, Almoalem_Evidence_2022}: $J_c \propto \Delta T_c^{3/2}$. This results in similar periodic scallops in $I_c$ with the flux threaded through the loop as for in $\Delta T_c$, as we do observe in our measurements.\\

For a more detailed derivation, the reader is referred to the book \textit{Introduction to Superconductivity} By Tinkham \cite{Tinkham}. Alternatively, the original 1964 publication by Little and Parks \cite{LittleParks_Fluxoid_1964} also provides a less thorough, but more intuitive derivation, which is directly linked to their experiments. Note: both Tinkham, and Little and Parks present their derivations in Gaussian cgs units, while we have chosen to work in SI units.\\

\section{Little-Parks oscillations - Supplementary data}
\label{Suppl:LP-data}
Seven devices with three different cross-sectional designs were measured. The oscillations in $I_c$ observed in the magnetotransport maps were isolated using a peak finder on each IV trace, for all magnetic field values. Then the slowly varying background was removed using a Savitsky-Golay filter, isolating just the oscillations in $I_c$. Now the periodicity in field can simply be determined through a FFT.

This analysis method is illustrated step-by-step on the next page, followed by an overview of the most important data of each loop device. Results are summarized in the table below.\\

\begin{table}[h]
    \centering
    \begin{tabular}{|c||c|c|c|c|}\hline
    \textbf{Device} & \textbf{Side length (design)} & \textbf{Corrected loop area} & \textbf{Expected $\Delta B$} & \textbf{Measured $\Delta B$ }\\
        & [nm] & [nm$^2$] & [mT] & [mT]  \\\hline\hline
    A   & 500 &  213.8 &  9.7 &   9.9 \\\hline
    B   & 250 &   36.5 & 56.7 &  49.4 \\\hline
    C   & 750 &  496.8 &  4.2 &   4.3 \\\hline
    D   & 750 &  512.5 &  4.0 &   4.3 \\\hline
    E   & 250 &   50.3 & 41.1 &  41.9 \\\hline
    F   & 500 &  215.1 &  9.6 &   9.8 \\\hline
    G   & 500 &  223.7 &  9.2 &  10.0 \\\hline
    \end{tabular}
\end{table}

\newpage\newpage
\begin{figure}[H]
\centering
\includegraphics[width=0.9\linewidth]{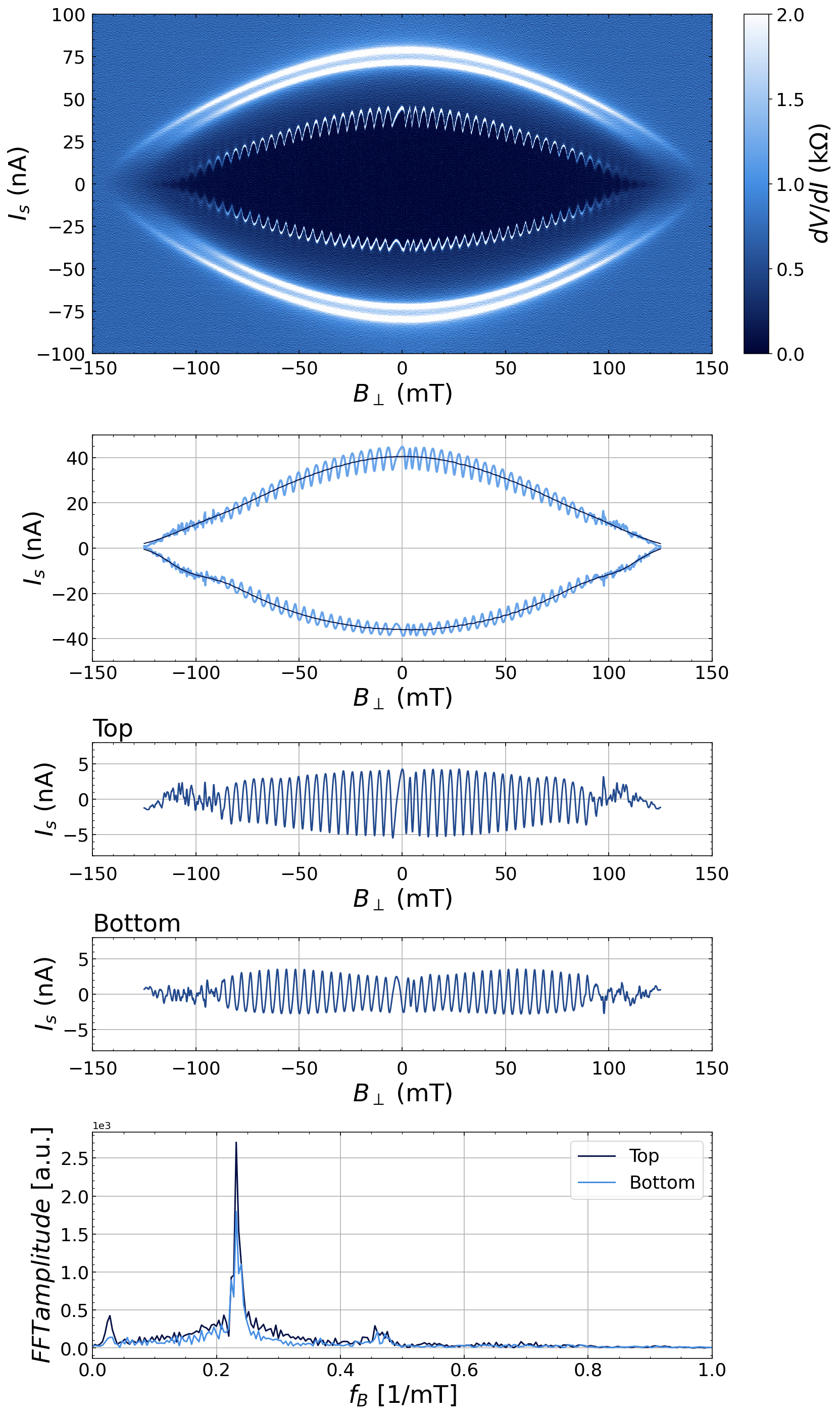}
\end{figure}

\begin{figure}[H]
\centering
\includegraphics[trim=20pt 20pt 20pt 20pt, clip, width=\linewidth]{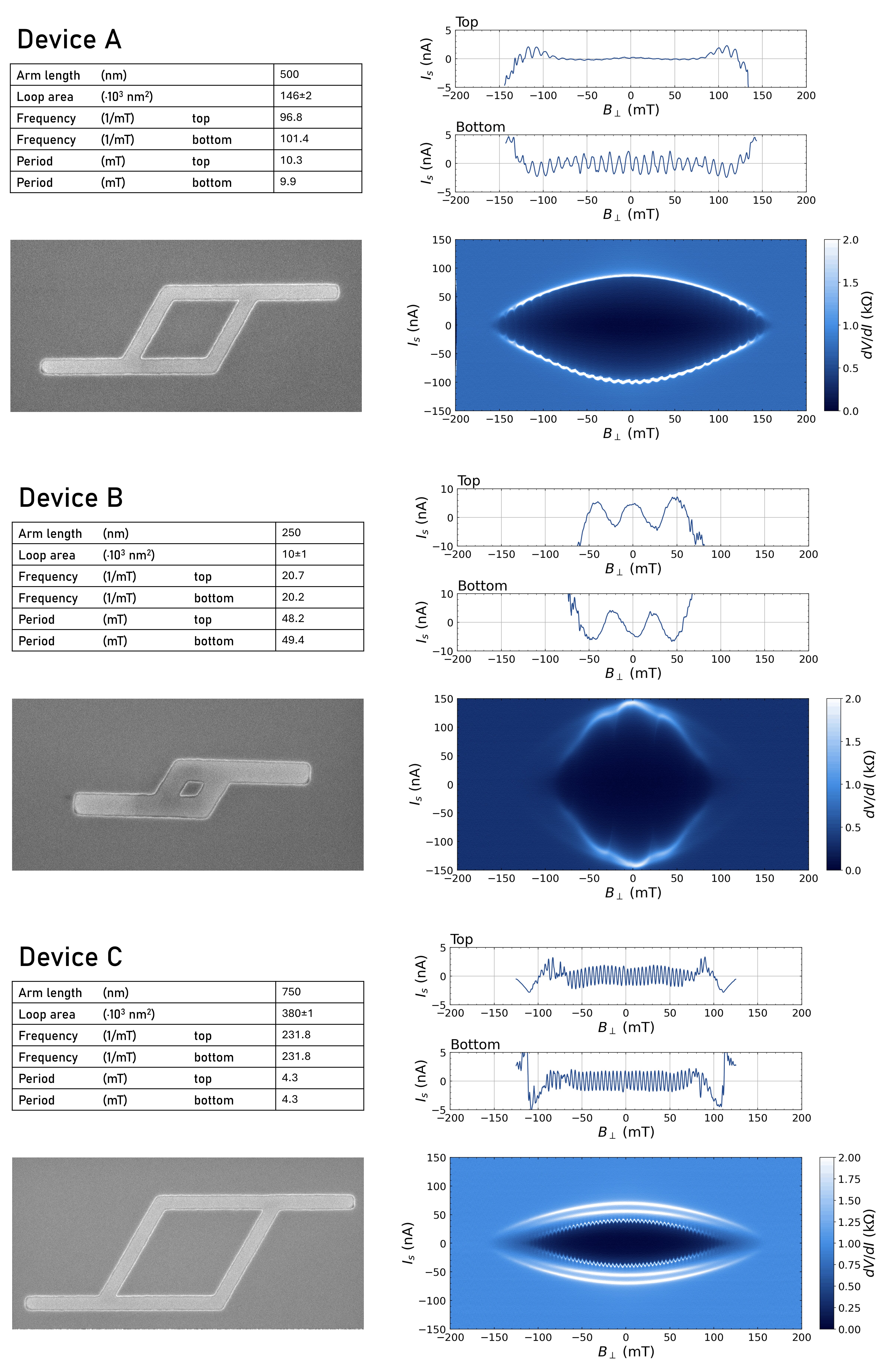}
\end{figure}

\begin{figure}[H]
\centering
\includegraphics[trim=20pt 20pt 20pt 20pt, clip, width=\linewidth]{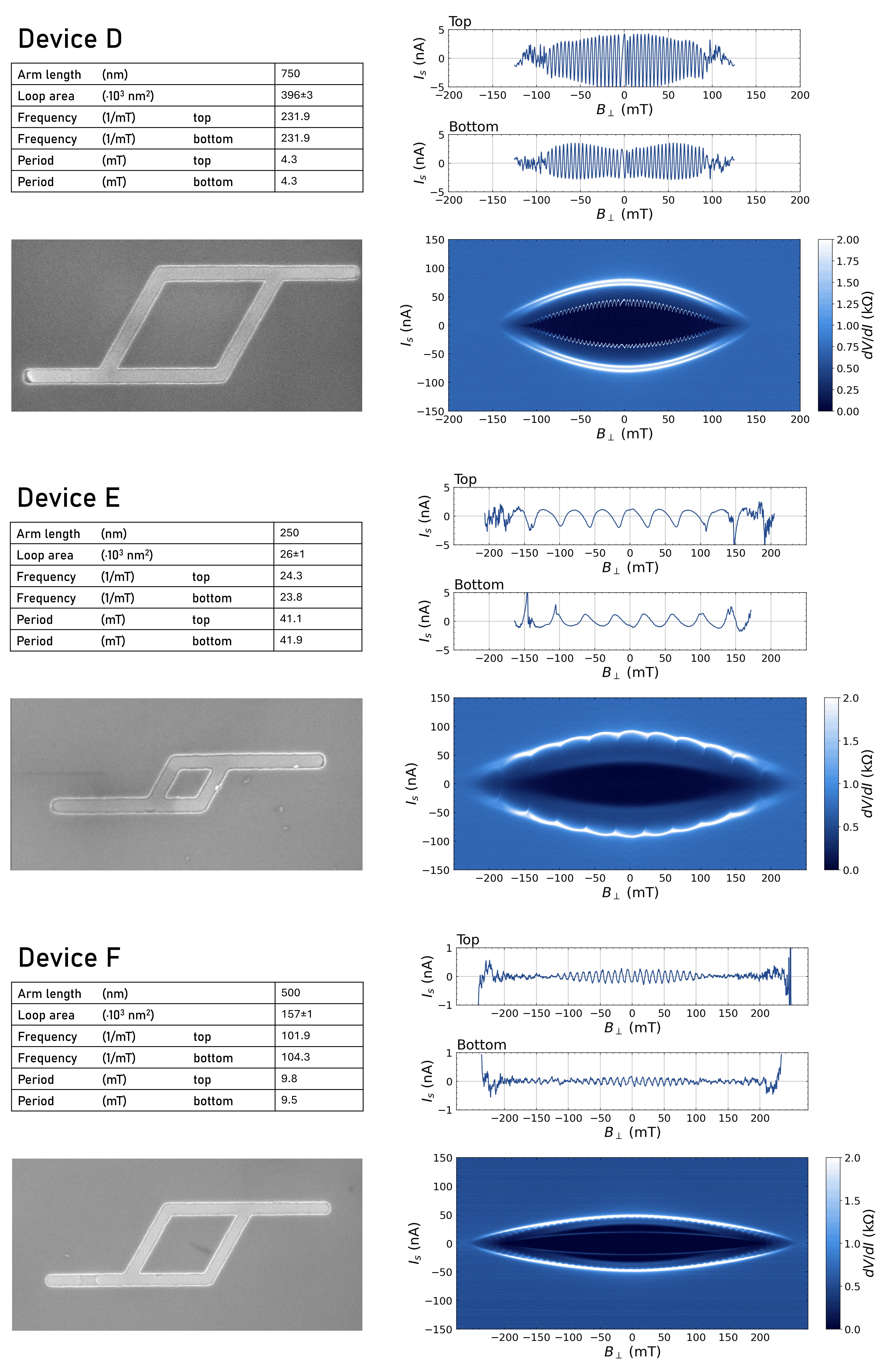}
\end{figure}

\begin{figure}[H]
\centering
\includegraphics[trim=20pt 20pt 20pt 20pt, clip, width=\linewidth]{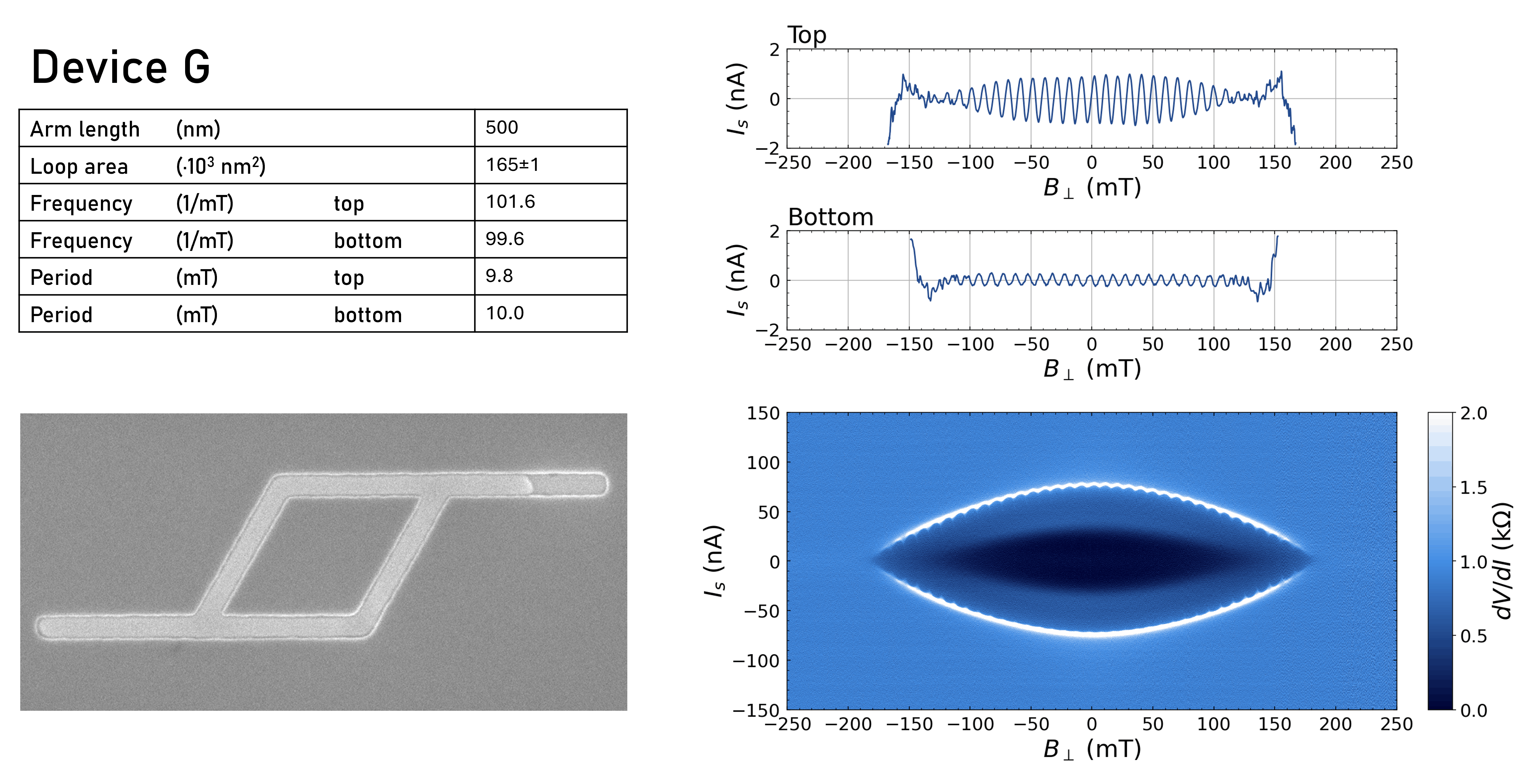}
\end{figure}

\end{document}